\documentclass[prl,aps,floats]{revtex4}
\usepackage{epsfig}
\begin{document}
\def\beq{\begin{equation}}
\def\eeq{\end{equation}}
\def\ber{\begin{eqnarray}}
\def\eer{\end{eqnarray}}
\def\l{\Lambda}
\def\b{{\rm b}}
\def\m{{\rm m}}
\def\om{\Omega_{0\rm m}}
\def\oml{\Omega_l}
\def\omx{\Omega_{\l_{\rm b}}}
\def \lleq {\lower0.9ex\hbox{ $\buildrel < \over \sim$} ~}
\def \ggeq {\lower0.9ex\hbox{ $\buildrel > \over \sim$} ~}
\def\rhoc{\rho_{0 {\rm c}}}

\def\etal{{\it et al.}}
\def\ie{i.e.~}
\def\n{\noindent}

\def\apj{Astroph.~J.~}
\def\mn{Mon.~Not.~Roy.~Ast.~Soc.~}
\def\asta{Astron.~Astrophys.~}
\def\aj{Astron.~J.~}
\def\prl{Phys.~Rev.~Lett.~}
\def\prd{Phys.~Rev.~D~}
\def\nucp{Nucl.~Phys.~}
\def\nat{Nature~}
\def\plb{Phys.~Lett.~B~}
\def\jetpl{JETP ~Lett.~}

\title{Exploring the Properties of Dark Energy Using Type Ia Supernovae and 
Other Datasets}

\author{Ujjaini Alam$^{a}$, Varun Sahni$^b$ and Alexei A. Starobinsky$^c$}
\affiliation{$^a$International Centre for Theoretical Physics, Strada Costiera 11,
34100 Trieste, Italy}
\affiliation{$^b$ Inter-University Centre for Astronomy and Astrophysics, 
Pune 411~007, India}
\affiliation{$^c$ Landau Institute for Theoretical Physics, 119334 Moscow, Russia}

\thispagestyle{empty}

\sloppy

\begin{abstract}
\small{ 
We reconstruct dark energy properties from two complementary supernova
datasets -- the newly released Gold+HST sample and SNLS. The results
obtained are consistent with standard $\l$CDM model within $2\sigma$
error bars although the Gold+HST data favour evolving dark energy
slightly more than SNLS. Using complementary data from baryon acoustic
oscillations and the cosmic microwave background to constrain dark
energy, we find that our results in this case are strongly dependent
on the present value of the matter density $\om$.  Consequently, no
firm conclusions regarding constancy or variability of dark energy
density can be drawn from these data alone unless the value of $\om$
is known to an accuracy of a few percent. However, possible
variability is significantly restricted if this data is used in
conjunction with supernova data.}
\end{abstract}

\maketitle

\section{Introduction}
\label{sec:intro}

There is a growing consensus in cosmology that the Universe is currently
accelerating. Perhaps the simplest explanation of this property is the 
presence of a positive cosmological constant $\Lambda$. Although $\Lambda$ 
appears to explain all current observations satisfactorily, to do so its value 
must necessarily be very small $\Lambda/8\pi G \simeq 10^{-47}$GeV$^4$. So,
it represents a new small constant of nature in addition to those known
from elementary particle physics. However, since it is not known at present
how to derive it from these other small constants and it is even unclear if it
should be exactly constant, other phenomenological explanations for cosmic 
acceleration have been suggested. Collectively called {\em dark energy} (DE)
models, they are based either on the introduction of new physical fields 
(quintessence and phantom models, the Chaplygin gas, etc.),
or on {\em geometrical approaches} which attempt to generate acceleration
by means of a change in the laws of gravity and, therefore, the geometry of 
space-time \cite{DE_review}. Scalar-tensor gravity, $R + f(R)$ gravity
and higher dimensional `Braneworld' models are prominent members of this
second category. 

The growing number of DE models has inspired a complementary approach
whose aim is to reconstruct properties of DE directly from
observations in a quasi-model independent manner, see the recent
review \cite{ss06} and the extensive list of references therein. The
aim of this paper is to investigate what new insights can be obtained
about DE using the most recent data, and to see whether or not these
data strengthen previously obtained results on the closeness of DE to
$\Lambda$. We study two different supernova samples -- the newly
released Gold+HST sample~ \cite{riess06} and SNLS data \cite{astier05}
-- to see what constraints follow from them on possible evolution of
DE. We also investigate the possibility of extracting information
about the nature of DE from datasets other than the supernova data. To
this end, we consider what follows from the value of the $R$ parameter
characterizing acoustic peaks in the angular power spectrum of the
cosmic microwave background (CMB)~ \cite{wmap,wang} and from the SDSS
result on the baryon acoustic oscillation peak
(BAO)~\cite{eisenstein05}.

\section{Data and Methodology}
\label{sec:method}

In this paper, we shall compare the reconstruction results for
different recent sets of observations. We briefly summarize each of
the data sets which we shall use before proceeding to give the results
of our comparison.

Type Ia supernova data is the strongest evidence for DE in current
cosmology.  The data is in the form of $\mu_{0,i}(z_i)=m_B-M=5{\rm
  log}d_L(z_i)+25$ with
\beq
d_L(z)=(1+z)\int_0^z \frac{dz}{H(z)}\,\,,
\eeq
where $H(z) = H_0 h(z)$ is defined in (\ref{eq:hpoly}) for the ansatz
used.  

We use the following two SNe datasets available to us at present.

\begin{enumerate}
  
\item 
{\em The Gold+HST data set} : As recently as 2003, the entire
supernova dataset from the two different surveys -- Supernova
Cosmology Project (SCP) and High $z$ Supernova Search Team (HZT),
along with low redshift supernovae from Calan-Tololo Supernova Search
(CTSS) comprised of a meager $92$ supernovae
\cite{perl98a,riess98,perl98b}, with very few at high redshifts,
$z>0.7$. The method of data reduction for the different teams was also
somewhat different, so that it was not possible to use the supernovae
from the two datasets concurrently. The picture changed somewhat
dramatically during 2003-2004, when a set of papers from both these
teams \cite{tonry03,knop03,barris04} presented a joint dataset of
$194$ supernovae which used the same data reduction method. This new
data resulted in doubling the dataset at $z>0.7$.  Not all these
supernovae could be identified beyond doubt as Type Ia supernovae
however, in many cases complete spectral data was not available. In
early 2004, Riess \etal~\cite{riess04} reanalyzed the data with
somewhat more rigorous standards, excluding several supernovae for
uncertain classification or inaccurate colour measurements. They also
added $14$ new high redshift supernovae observed by the Hubble Space
Telescope (HST) to this sample. This resulted in a sample known as the
'Gold' dataset. The latest publication from the same team
\cite{riess06} adds a further $10$ SNe from HST to the dataset, and
excludes data below $cz = 7000 km/s \ (z=0.0233)$ to avoid the
influence of a possible local ``Hubble Bubble''. The final sample now
comprises of $135$ supernovae (the furthest being at redshift
$z=1.755$) -- we call this sample the `Gold+HST' dataset and use it as
our first SNe sample.

\item
{\em The Supernova Legacy Survey SNe data set (SNLS)} : The SuperNova Legacy
Survey \cite{snls} is an ongoing 5-year project which is expected to
yield more than $700$ spectroscopically confirmed supernovae below
redshift of one. The first two-year results from this survey
\cite{astier05} have provided us with 71 new supernovae below $z=1$.
We shall use these 71 SNe together with the already available low-$z$
supernova data, \ie  a total of $115$ SNe, as our second supernova
sample.
  
\end{enumerate}

In \cite{riess06}, the SNLS and Gold datasets have also been merged
together for some of the results. Since the standardization techniques
of the two teams are somewhat different, as of now, we consider the
two samples separately till a better handle on the systematics is
obtained.

Although supernova data forms the major observational proof of dark
energy, we may try to obtain independent information on the nature of dark
energy from the other observational results. Here we look at the
following two datasets.

\begin{enumerate}

\item
Observations of the cosmic microwave background provide us with very
accurate measurements, which may be used to gain insight about dark
energy \cite{wmap}. We may use the WMAP 3-year results to get~\cite{wang}
\beq
R = \sqrt{\om}\int_0^{z_{\rm ls}} \frac{dz}{h(z)} = 1.70 \pm 0.03
\eeq 
where $h(z) = H(z)/H_0$ is defined in (\ref{eq:hpoly}) for the ansatz
used. For calculating this quantity, we use $\Omega_b h^2=0.024$ and
$\om h^2 =0.14 \pm 0.02$. To calculate $z_{ls}$ we
use a fitting function given in \cite{hu}:
\beq
z_{ls}=1048 [1+0.00124 (\Omega_b h^2)^{-0.738}][1+g_1 (\om h^2)^{g_2}]\,\,,
\eeq
where the quantities $g_1,g_2$ are defined as
\ber
g_1 &=& 0.078(\Omega_b h^2)^{-0.238} [1+39.5 (\Omega_b h^2)^{0.763}]^{-1},\\
g_2 &=& 0.56 [1+21.1 (\Omega_b h^2)^{1.81}]^{-1}\,\,.
\eer

\item 
{\em The Baryon Acoustic Oscillation Peak} : A remarkable confirmation
of the standard big bang cosmology has been the recent detection of a
peak in the correlation function of luminous red galaxies in the Sloan
Digital Sky Survey \cite{eisenstein05}.  This peak, which is predicted
to arise precisely at the measured scale of $100$ h$^{-1}$ Mpc due to
acoustic oscillations in the primordial baryon-photon plasma prior to
recombination, can provide a `standard ruler' with which to test dark
energy models. Specifically, we shall use the value
\cite{eisenstein05}
\beq
A = \frac{\sqrt{\om}}{h(z_1)^{1/3}}~\bigg\lbrack ~\frac{1}{z_1}~\int_0^{z_1}\frac{dz}{h(z)}
~\bigg\rbrack^{2/3} = ~0.469 \left(\frac{n}{0.98}\right)^{-0.35} \pm 0.017~,
\eeq
where $h(z) = H(z)/H_0$ is defined in (\ref{eq:hpoly}) for the ansatz
used, and $z_1 = 0.35$ is the redshift at which the acoustic scale has
been measured.  The 3-year WMAP results yield $n=0.95$ for
the spectral index of density perturbations.

\end{enumerate}

We note here that the quantity $A$ measured from the baryon acoustic
oscillations appears to be independent of the value of $h$. However,
the quantity $R$ obtained from CMB measurements depends on $\om h^2$,
and therefore on the Hubble parameter. Therefore, when using this
quantity, we require to either assign some value to $h$ or to
marginalize over $\om h^2$.

We use $\chi^2$ minimization on the different datasets with 
\beq
\chi^2(H_0,p_j)=\sum_i \frac{[y_{{\rm fit},i}(z_i;H_0,\om,p_j)-y_{i}]^2}{\sigma^2_i}\,\,.
\eeq
Here, $y_i$ is the data at redshift of $z_i$ and $\sigma_i$ is the
uncertainty in the individual $y_i$, and $p_j$ are the parameters
$\om,A_1,A_2$ for the ansatz \cite{state}
\beq\label{eq:hpoly}
h^2(z)=\frac{H^2(z)}{H_0^2}=\om x^3+A_0+A_1x+A_2 x^2\,\,, ~~x=1+z
\eeq
with $A_0=1-\om-A_1-A_2$ for a flat universe. Note that
this ansatz for the Hubble parameter has been tested and found to give
accurate results for a variety of cosmological models
\cite{alam04a,alam04b}. Its main features are: (i) it accounts for a
matter dominated regime at high redshift, (ii) it is exact for DE
being the mixture of a cosmological constant and non-relativistic domain walls
and cosmic strings (in other words, for DE consisting of 3 components with
with $w=-1$, $w=-1/3$ and $w=-2/3$),
(iii) it can satisfactorily emulate DE with an evolving equation of state 
including Braneworld models and phantom models admitting regions with 
$w < -1$ (moreover, the DE effective energy density $\rho_{DE}$ may even 
become negative in such models).

\section{Results}

To test the two SNe datasets, we first reconstruct a $\l$CDM model
with curvature, using the formula
\beq
d_L(z)=\frac{c(1+z)}{\sqrt{|\kappa|}} {\cal S} \left ( \sqrt{|\kappa|} \int_0^z \left [(1+z^{\prime})^2(1+\Omega_{\m} z^{\prime})-z^{\prime}(2+z^{\prime})\Omega_{\l} \right]^{-1/2} dz^{\prime} \right) \,\,,
\eeq
where
\ber
&&\Omega_{\m}+\Omega_{\l} > 1 \Rightarrow {\cal S}(x)={\rm sin}(x), \ \kappa=1-\Omega_{\m}-\Omega_{\l} \nonumber\\
&&\Omega_{\m}+\Omega_{\l} < 1 \Rightarrow {\cal S}(x)={\rm sinh}(x), \ \kappa=1-\Omega_{\m}-\Omega_{\l} \nonumber\\
&&\Omega_{\m}+\Omega_{\l} = 1 \Rightarrow {\cal S}(x)=x, \ \kappa=1 \nonumber\,\,.
\eer

\begin{figure*} 
\centering
\begin{center}
$\begin{array}{@{\hspace{-1.0in}}c@{\hspace{0.0in}}c}
\multicolumn{1}{l}{\mbox{}} &
\multicolumn{1}{l}{\mbox{}} \\ [-0.20in]
\epsfxsize=3.8in
\epsffile{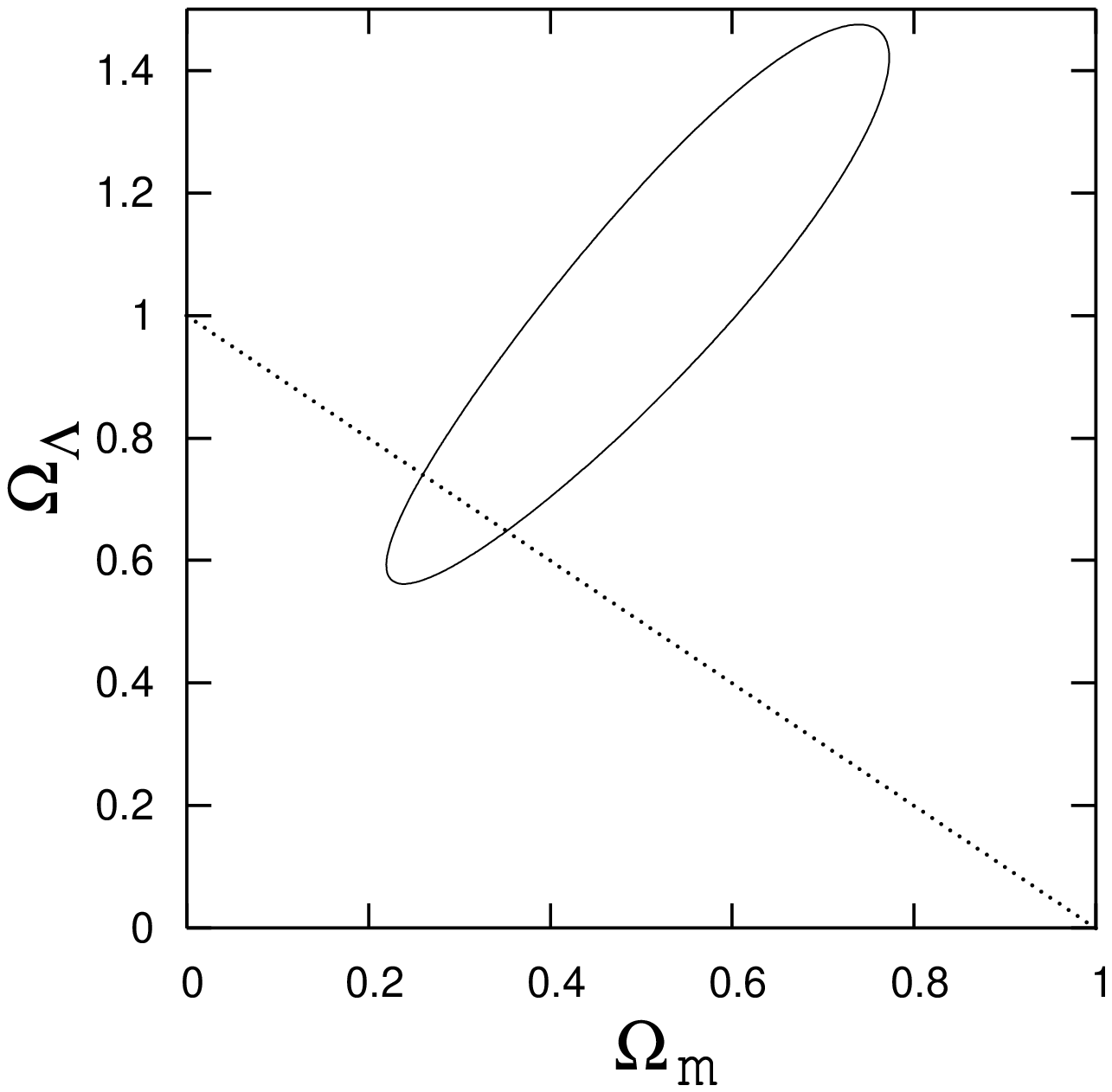} &  
\epsfxsize=3.8in
\epsffile{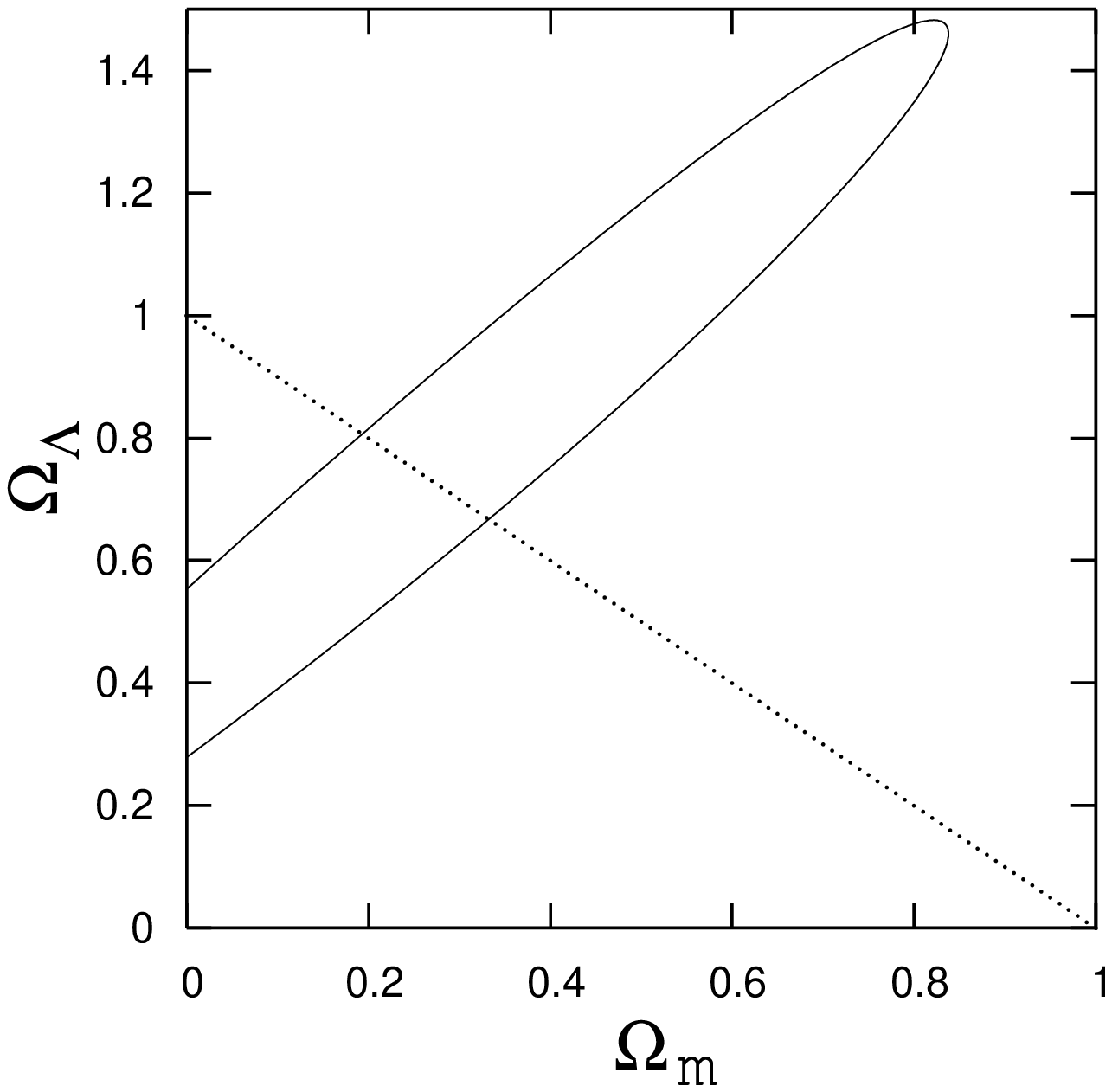} \\
\end{array}$
\end{center}
\vspace{0.0cm}
\caption{\small 
$2\sigma$ confidence levels are shown for the Gold+HST dataset (left
panel) and SNLS dataset (right panel) in the $\om,\Omega_{\l}$ plane,
with the dotted line showing the flat universe. }
\label{fig:cc}
\end{figure*}

In figure~\ref{fig:cc} we show results for a $\l$CDM model for both
the Gold+HST and SNLS datasets. We see that for the Gold+HST data, the
dataset prefers a closed universe somewhat over the flat model. The
SNLS best-fit, on the other hand, is very close to the flat model.

\begin{figure*} 
\centering
\begin{center}
$\begin{array}{@{\hspace{-1.0in}}c@{\hspace{0.0in}}c}
\multicolumn{1}{l}{\mbox{}} &
\multicolumn{1}{l}{\mbox{}} \\ [-0.20in]
\epsfxsize=3.8in
\epsffile{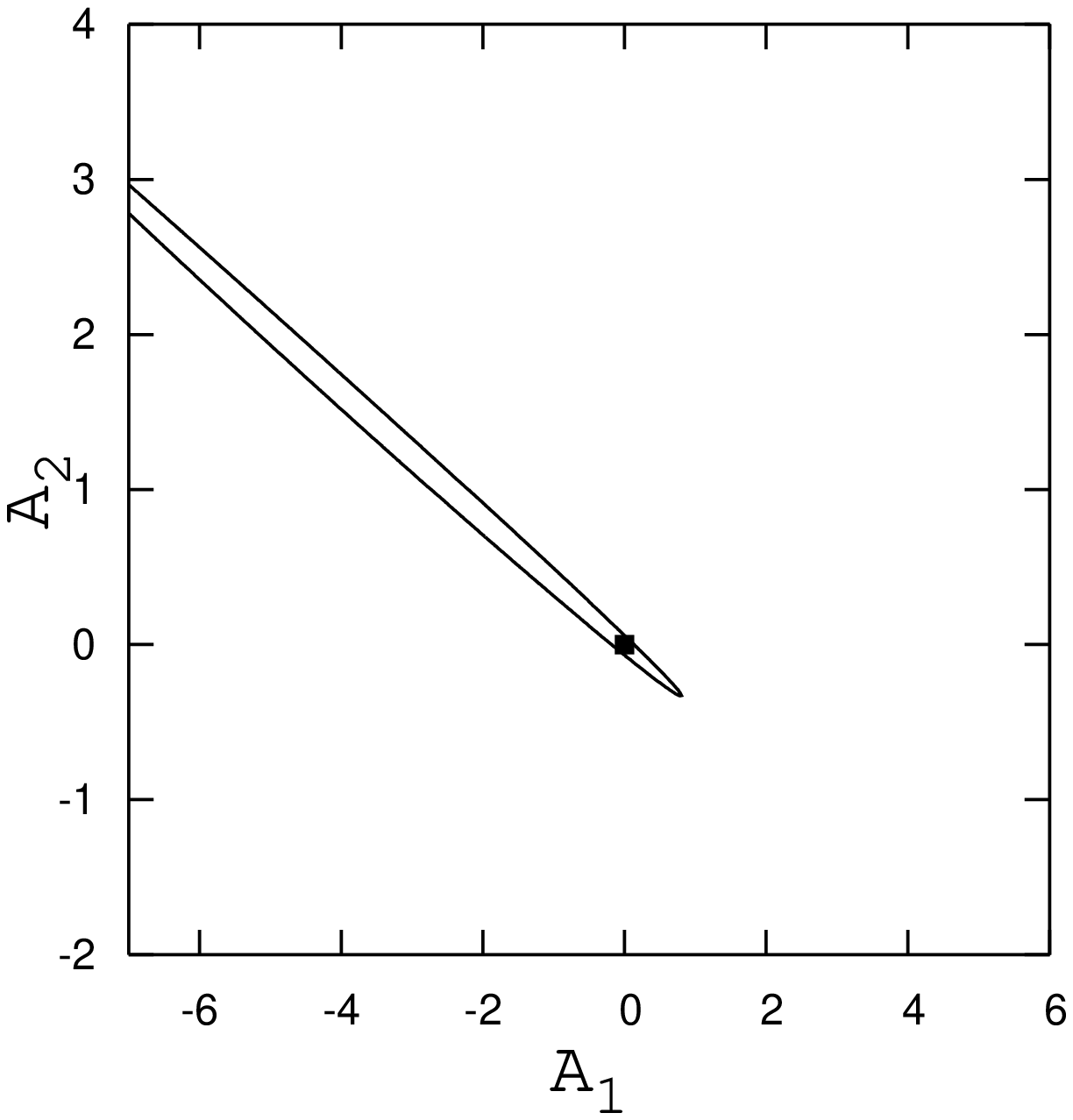} &  
\epsfxsize=3.in
\epsffile{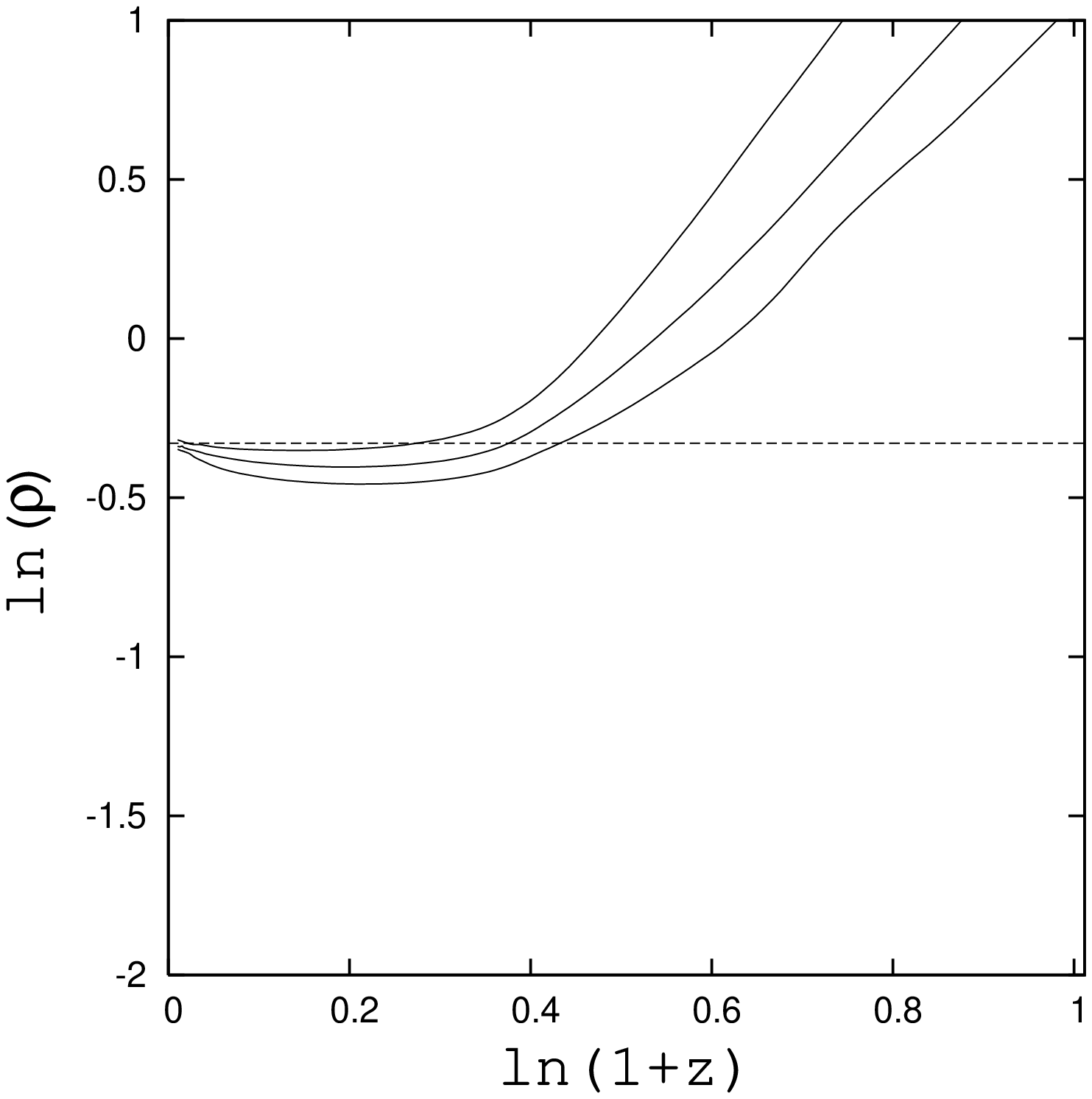} \\
\epsfxsize=3.in
\epsffile{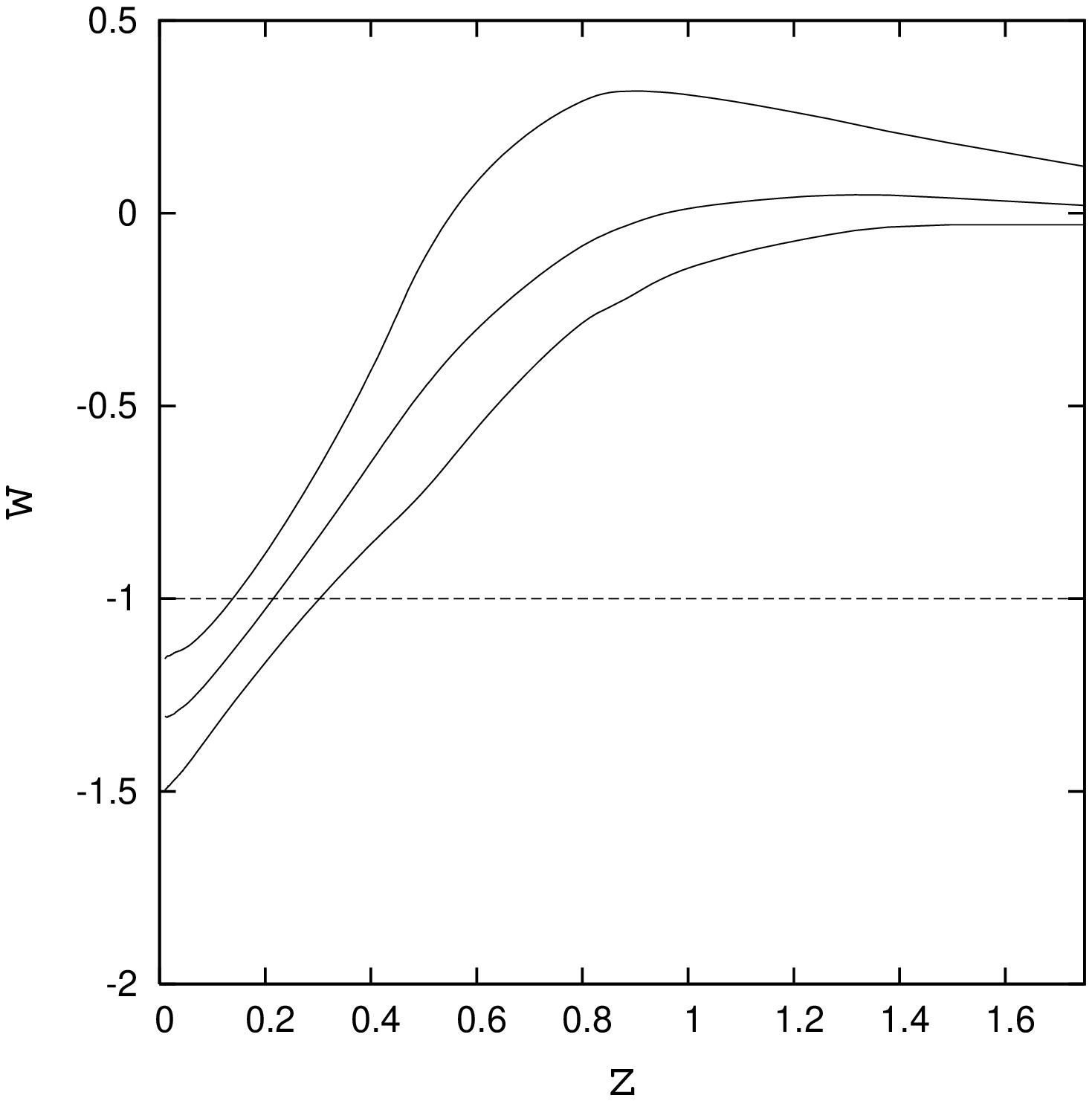} &
\epsfxsize=3.in
\epsffile{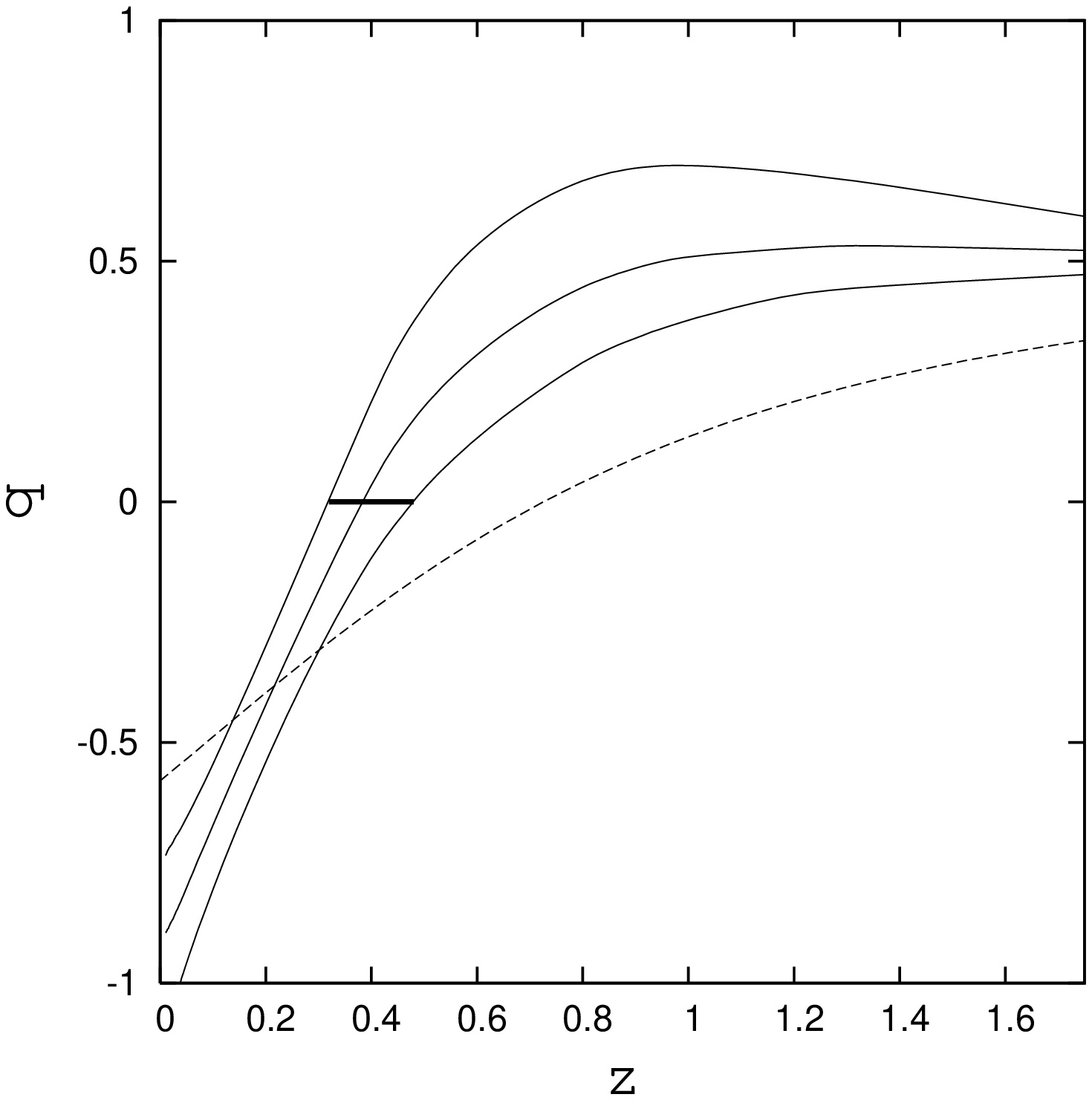} \\
\end{array}$
\end{center}
\vspace{0.0cm}
\caption{\small 
 $2\sigma$ confidence levels for the Gold dataset using
$\om=0.28\pm0.03$. The upper left hand panel shows the confidence
levels in $A_1-A_2$, with the black dot representing $\l$CDM. The
upper right hand panel shows the logarithmic $2\sigma$ variation of
the DE density in terms of redshift. The dashed line
represents $\l$CDM. The lower left and right hand panels represent the
variation of the equation of state and deceleration parameter
respectively. The dashed lines in both panels represent $\l$CDM. The thick
solid line in the lower right hand panel shows the acceleration epoch,
\ie the redshift at which the universe started accelerating.}
\label{fig:gold}
\end{figure*}

We next study the two datasets in more detail using
ansatz~(\ref{eq:hpoly}).  In figure~\ref{fig:gold} we show the results
for the Gold+HST SNe dataset. We marginalize over $\om=0.28\pm0.03$
which is the currently accepted value for matter density from SDSS
\cite{eisenstein05} and show the $2\sigma$ confidence levels in
$A_1-A_2$ in panel (a). The logarithmic variation of
$\rho_{\rm DE} = A_0+A_1x+A_2x^2$ is shown in panel (b). The
variation of the equation of state of DE
\beq\label{eq:state}
w(x) = 
\frac{(2 x /3) \ d \ {\rm ln}H \ / \ dx - 1}{1 \ - \ (H_0/H)^2
\Omega_{0m} \ x^3}\,\,
\eeq
is shown in panel (c), and the
deceleration parameter
\beq
q\equiv -\ddot a/aH^2 = \frac{H'(x)}{H(x)}~ x - 1 \, , ~~~ x = 1 + z \,
\label{eq:decel1}
\eeq
is plotted in panel (d). We find that the best-fit is somewhat away
from $\l$CDM, and favours an evolving model of DE, although
$\l$CDM is still within $2\sigma$ of the best-fit. From the
deceleration parameter we find the value of the acceleration epoch,
which is the redshift at which the universe started accelerating. For
the Gold data this transition between deceleration and acceleration
occurs between $z_a=0.32-0.48$ at $2\sigma$. The earlier Gold dataset
of \cite{riess04} favoured an evolving DE model starting from
$w < -1$ at present and evolving to $w \sim 0$ at $z \sim 1$ (as shown
in \cite{alam04c} and numerous subsequent papers), the addition of the 
10 new HST data points do not appear to change that result. These results 
also agree with the analysis shown in figure 12 of \cite{riess06}, as well 
as results from the non-parametric approach advocated in \cite{hut}.

\begin{figure*} 
\centering
\begin{center}
$\begin{array}{@{\hspace{-1.0in}}c@{\hspace{0.0in}}c}
\multicolumn{1}{l}{\mbox{}} &
\multicolumn{1}{l}{\mbox{}} \\ [-0.20in]
\epsfxsize=3.8in
\epsffile{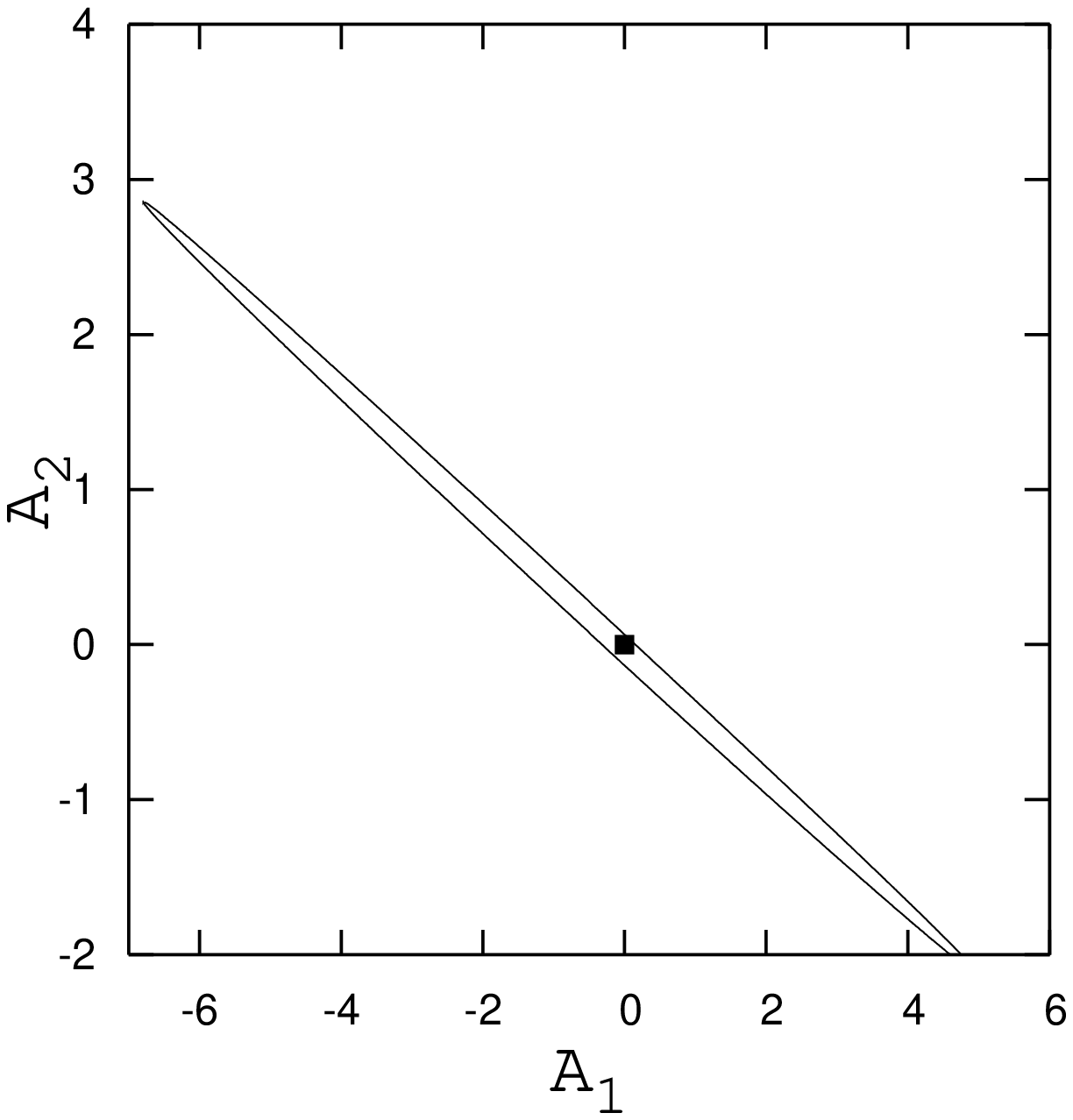} &  
\epsfxsize=3.in
\epsffile{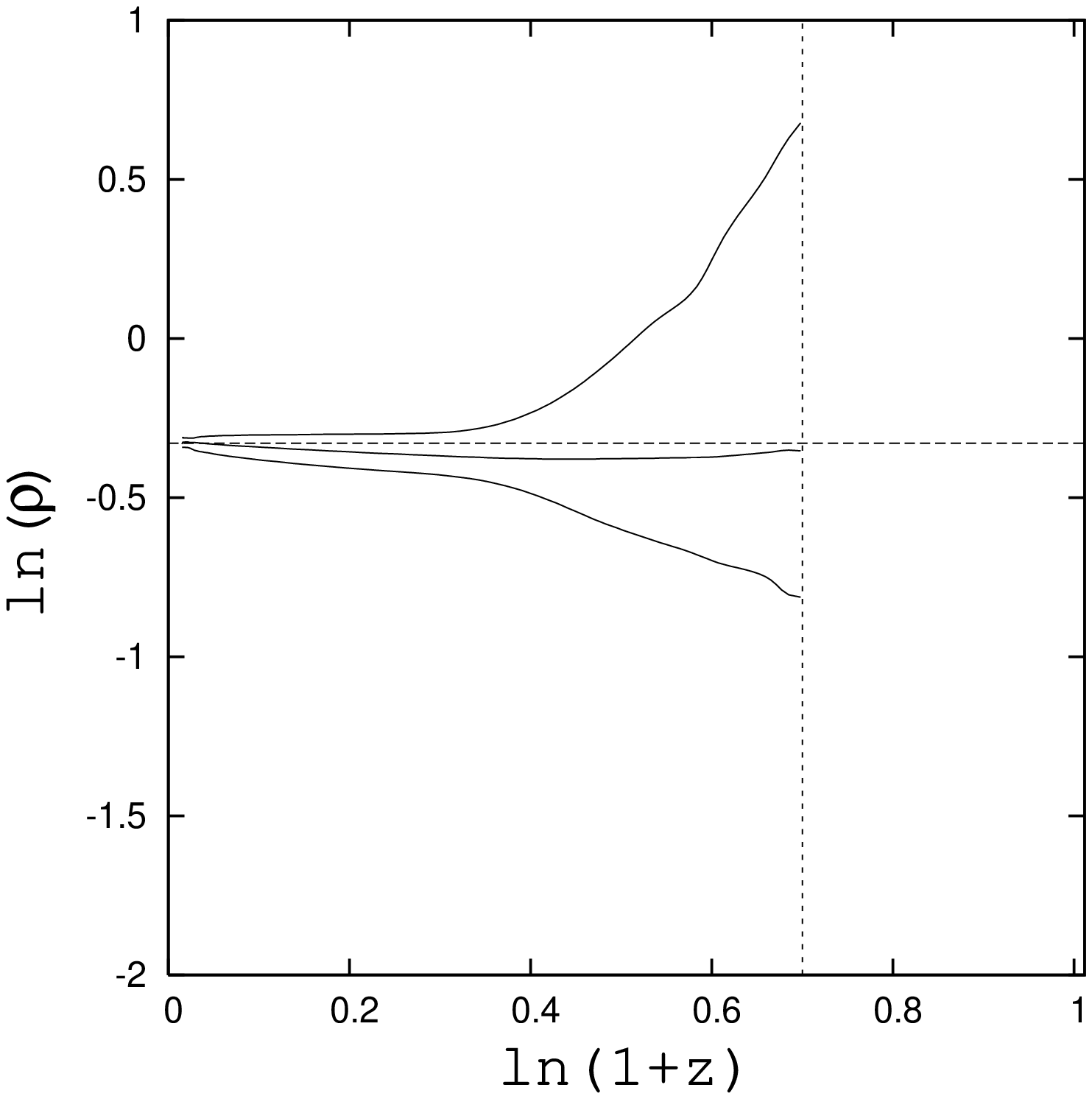} \\
\epsfxsize=3.in
\epsffile{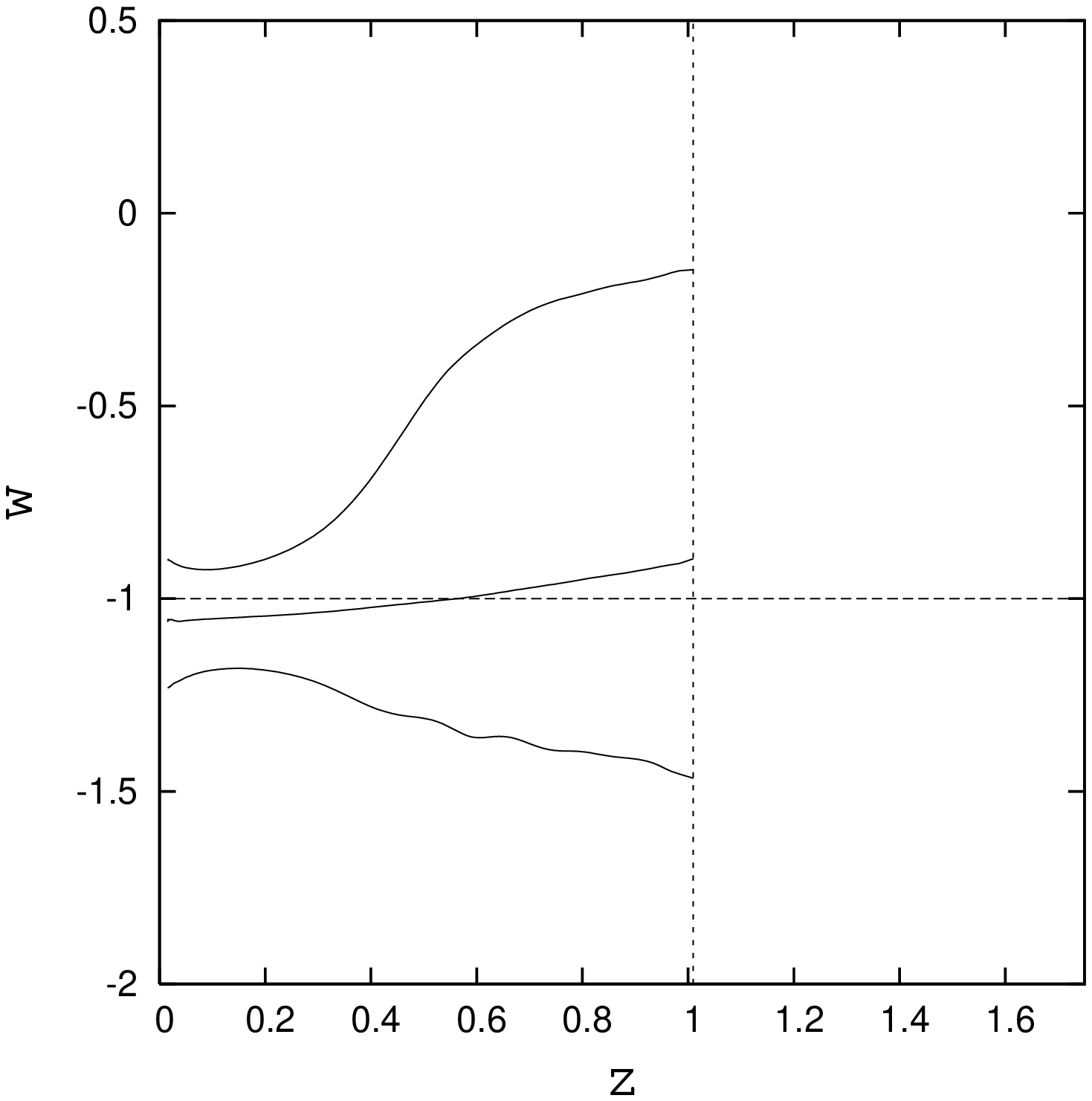} &
\epsfxsize=3.in
\epsffile{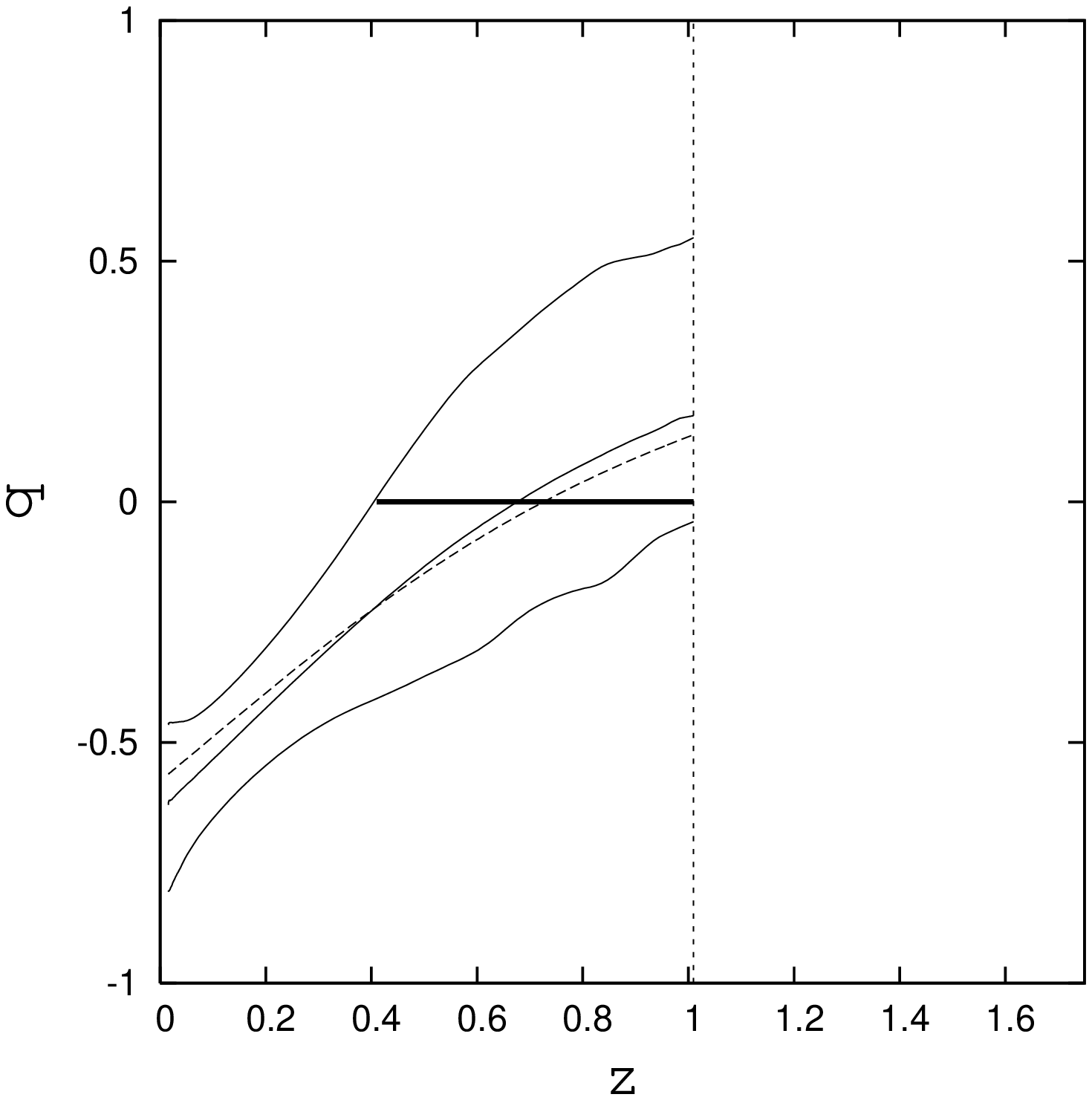} \\
\end{array}$
\end{center}
\vspace{0.0cm}
\caption{\small 
 $2\sigma$ confidence levels for the SNLS dataset using
$\om=0.28\pm0.03$. The upper left hand panel shows the confidence
levels in $A_1-A_2$, with the black dot representing $\l$CDM. The
upper right hand panel shows the logarithmic $2\sigma$ variation of
the DE density in terms of redshift. The dashed line
represents $\l$CDM. The lower left and right hand panels represent the
variation of the equation of state and deceleration parameter
respectively. The dashed lines in both panels represent $\l$CDM. The thick
solid line in the lower right hand panel shows the acceleration epoch,
\ie the redshift at which the universe started accelerating. Results
are shown upto $z=1.01$}
\label{fig:snls}

\end{figure*}

Figure~\ref{fig:snls} shows the results for the same
reconstruction for the SNLS dataset. Once again we marginalize over
the matter density. The results in this case are much closer to
$\l$CDM, the DE equation of state shows very mild
evolution. The acceleration epoch is in the range $z_a=0.41-1.0$.

We note here that although we have marginalized over a particular
matter density, results from the supernova data are only weakly
dependent on $\om$. Within the reasonable range $\om=0.22-0.34$ (the
$2\sigma$ limits on the current observations for $\om$), the results
do not change significantly.

The ansatz (\ref{eq:hpoly}) is useful not only for reconstructing the
DE density but also for determining an important related quantity -- the 
$w$-probe \cite{arman}. As its name suggests, the $w$-probe
provides us with important insights about the equation of
state. However, unlike $w(z)$ which is determined by differentiating
$H(z)$, the $w$-probe is related to the {\em difference} in the value
of the Hubble parameter and is defined as follows
\beq
1+\bar{w} = \frac{1}{\Delta \ {\rm ln}(1+z)} \int [1+w(z)] d \ {\rm ln}(1+z) 
= \frac{1}{3} \frac{\Delta \ {\rm ln}(\rho_{\rm DE}/\rhoc)}{\Delta \ {\rm ln}
(1+z)} \,\,.
\eeq
Here $\Delta$ denotes the total change in a variable between
integration limits.  It is important to note that the $w$-probe, when
studied over different redshift bins, can give us evidence for the
variation of the equation of state from the first derivative of the
data alone, \ie from the DE density.  (As demonstrated in
\cite{arman}, $\bar{w}$ is much less sensitive to uncertainties in the
value of the matter density than $w(z)$.)  Table~\ref{tab:wprobe}
shows the value of the $w$-probe in three redshift bins (approximately
equal in log$(1+z)$) for both Gold+HST and SNLS datasets. We see that
for the gold dataset $\bar{w}$ is close to $-1$ in the low redshift
bin, but in the two higher redshift bins $\bar{w}$ is closer to
zero. Thus the Gold+HST dataset seems to provide support for the
evolution of DE.  On the other hand, for the SNLS data, for
both redshift bins (upto redshift of unity), $\bar{w}$ is close to
$-1$, showing that this dataset favours the cosmological constant more
than Gold+HST.

In a way, the SNLS results contradict the results obtained for the
Gold dataset. This fact had been noted for an earlier Gold dataset in
\cite{leandros}. However, the difference is not very large, at
$2\sigma$ the results from both datasets are consistent with each
other. We should also keep in mind the fact that the light-curve
standardization of the two datasets is done using different methods,
MLCS2k2 for Gold+HST and SALT for SNLS. As shown in \cite{astier05},
different standardization techniques may lead to differences of upto
$0.16$ magnitude in the data. The discrepancy in cosmological results
may therefore be attributed partly to the different standardization
techniques. There are also possible effects from other sources in the
data, such as systematic noise and K-correction. Therefore it seems that
using just the two current SNe datasets, a firm conclusion cannot yet
be reached about the nature of DE. We may conclude that while the
cosmological constant is more or less consistent with both datasets,
evolving DE is still not ruled out.

\begin{figure*} 
\centering
\begin{center}
$\begin{array}{@{\hspace{-1.0in}}c@{\hspace{0.0in}}c}
\multicolumn{1}{l}{\mbox{}} &
\multicolumn{1}{l}{\mbox{}} \\ [-0.20in]
\epsfxsize=3.8in
\epsffile{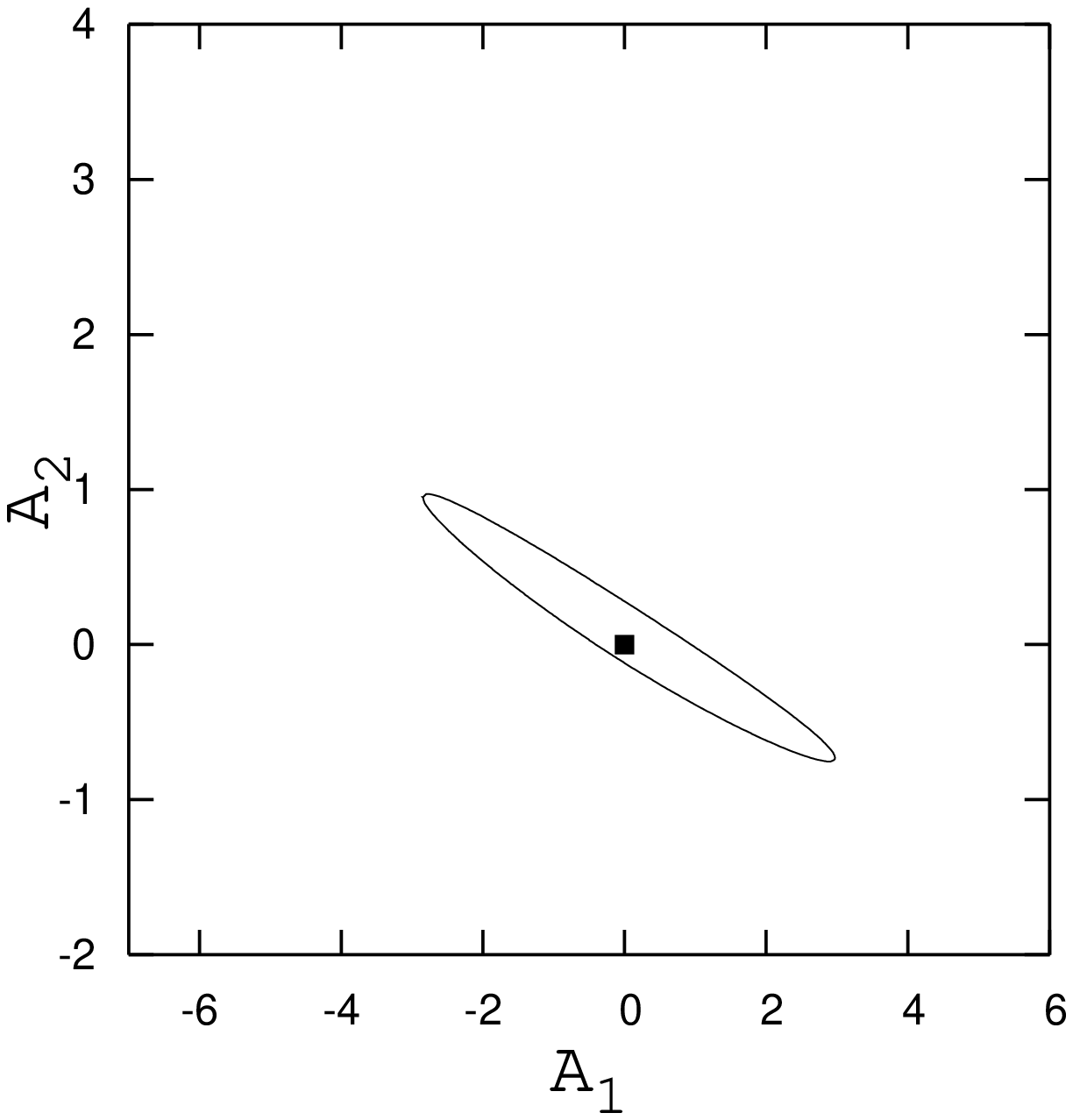} &  
\epsfxsize=3.in
\epsffile{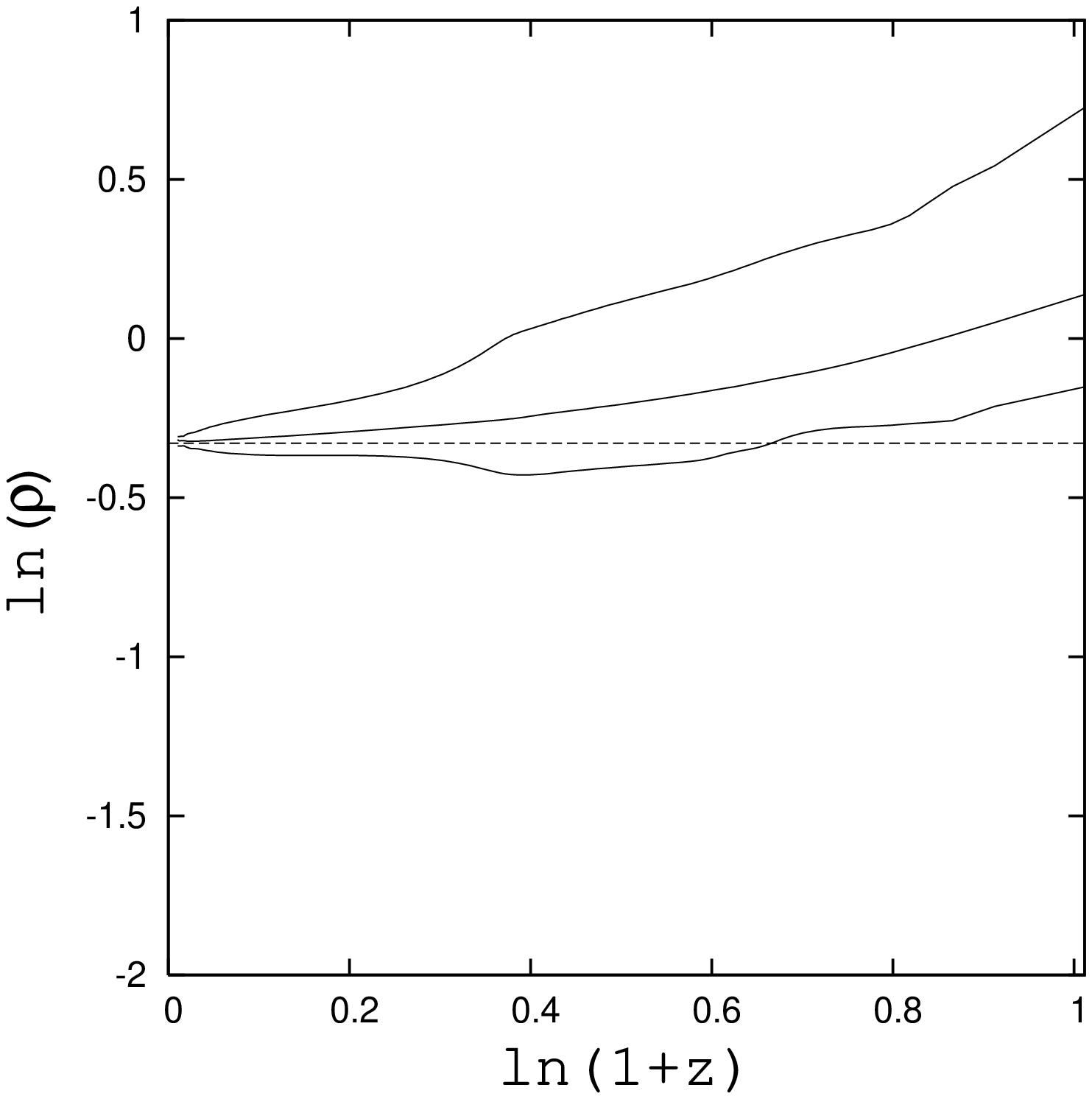} \\
\epsfxsize=3.in
\epsffile{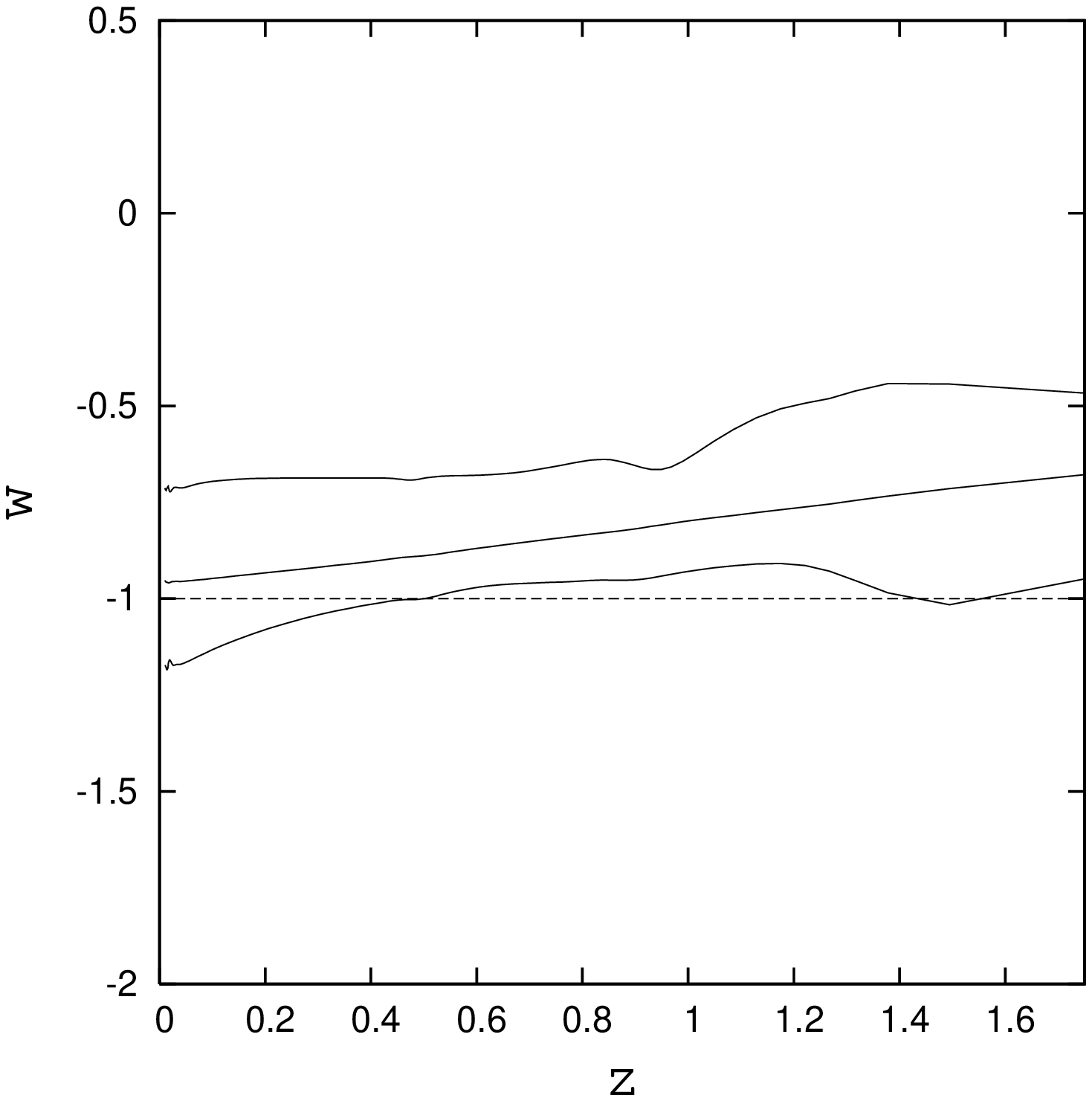} &
\epsfxsize=3.in
\epsffile{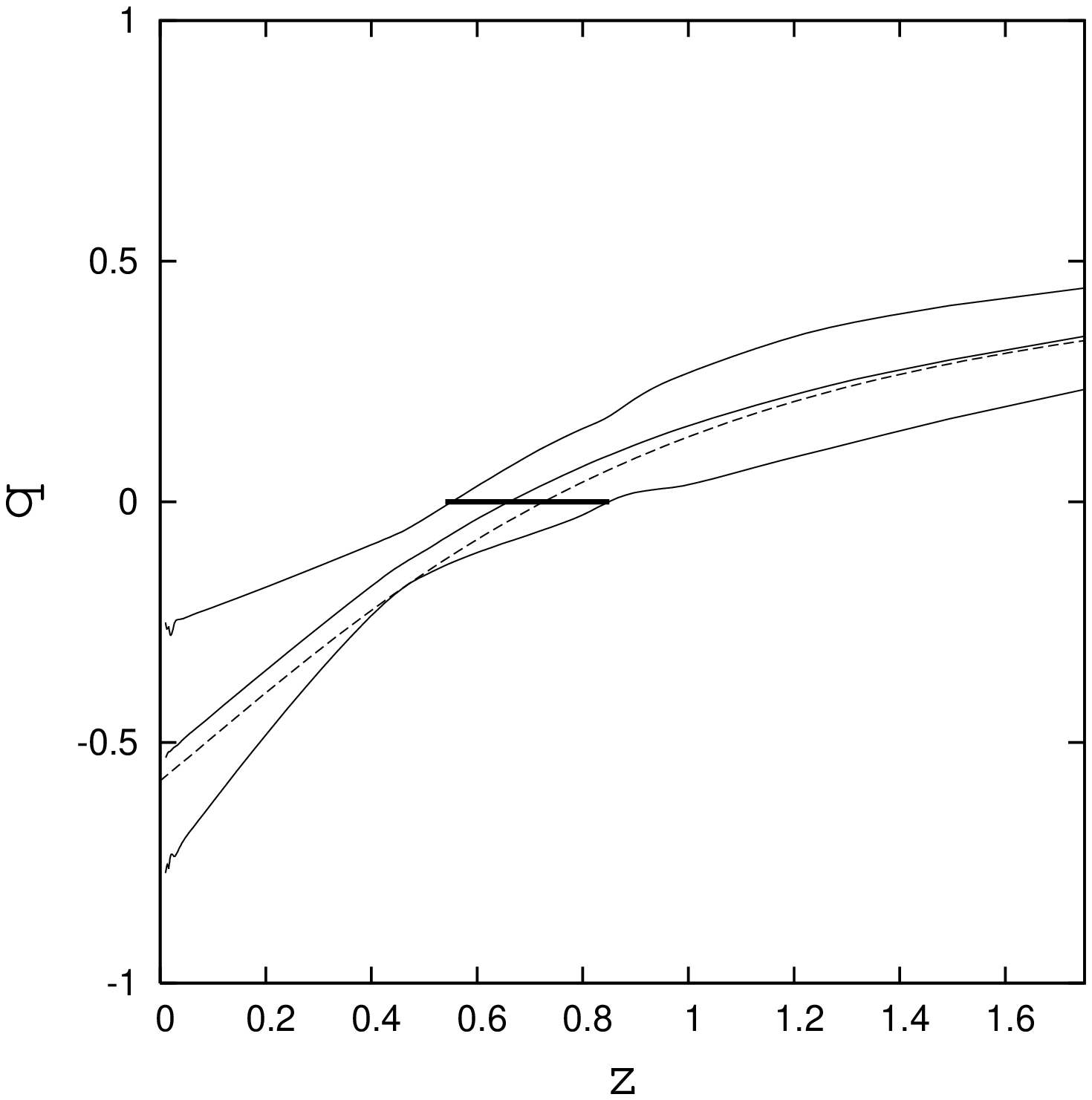} \\
\end{array}$
\end{center}
\vspace{0.0cm}
\caption{\small 
$2\sigma$ confidence levels for the CMB+BAO dataset using
$\om=0.28\pm0.03$. The upper left hand panel shows the confidence
levels in $A_1-A_2$, with the black dot representing $\l$CDM. The
upper right hand panel shows the logarithmic $2\sigma$ variation of
the DE density in terms of redshift. The dashed line
represents $\l$CDM. The lower left and right hand panels represent the
variation of the equation of state and deceleration parameter
respectively. The dashed lines in both panels represent $\l$CDM. The thick
solid line in the lower right hand panel shows the acceleration epoch,
\ie the redshift at which the universe started accelerating.}
\label{fig:cmb_sdss}
\end{figure*}

\begin{figure*} 
\centering
\begin{center}
$\begin{array}{@{\hspace{-1.0in}}c@{\hspace{-0.5in}}c@{\hspace{-0.5in}}c}
\multicolumn{1}{l}{\mbox{}} &
\multicolumn{1}{l}{\mbox{}} &
\multicolumn{1}{l}{\mbox{}} \\ [-0.20in]
\epsfxsize=3.in
\epsffile{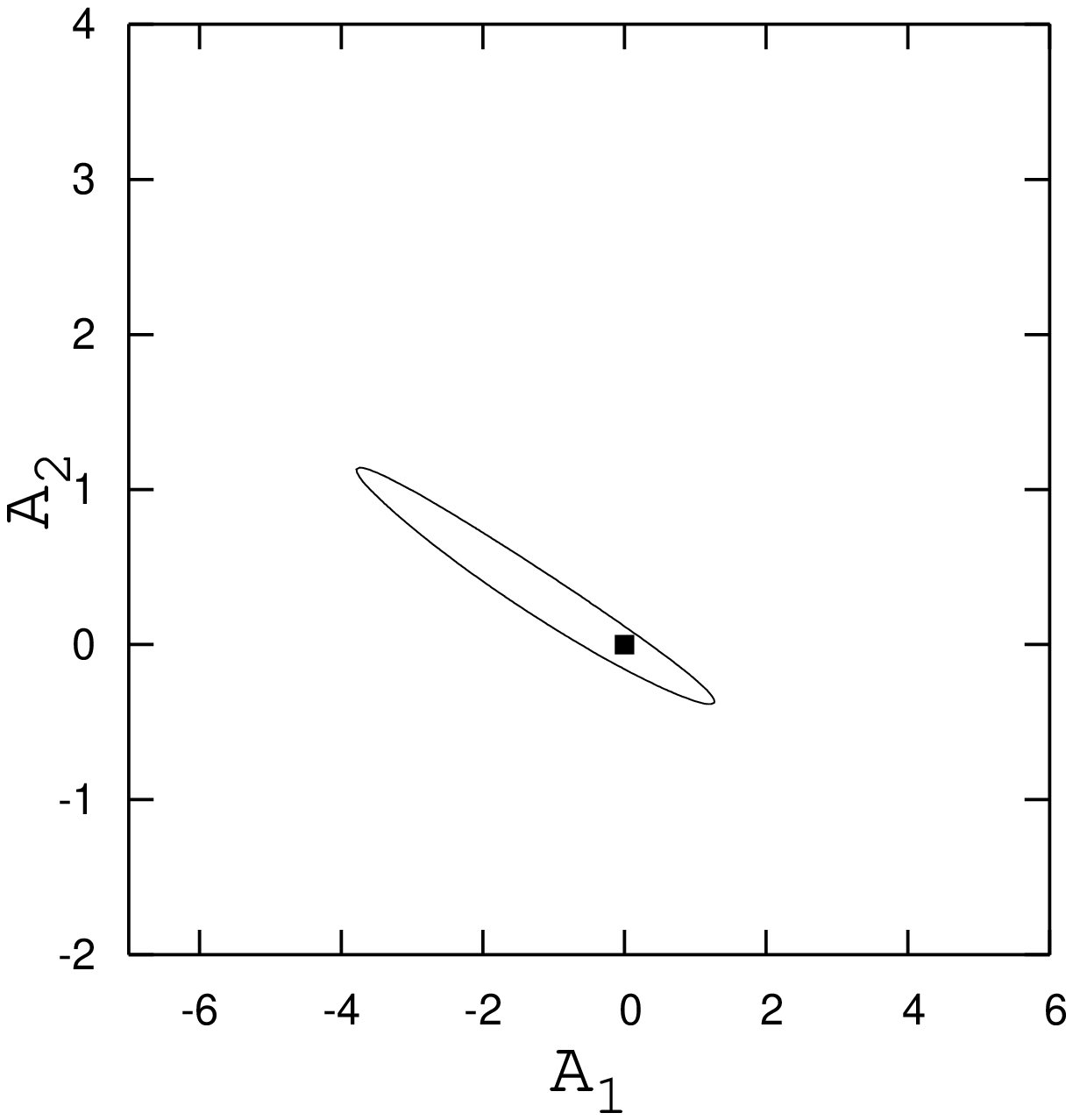} &  
\epsfxsize=3.in
\epsffile{conf_cmb_sdss_om=28.ps} &
\epsfxsize=3.in
\epsffile{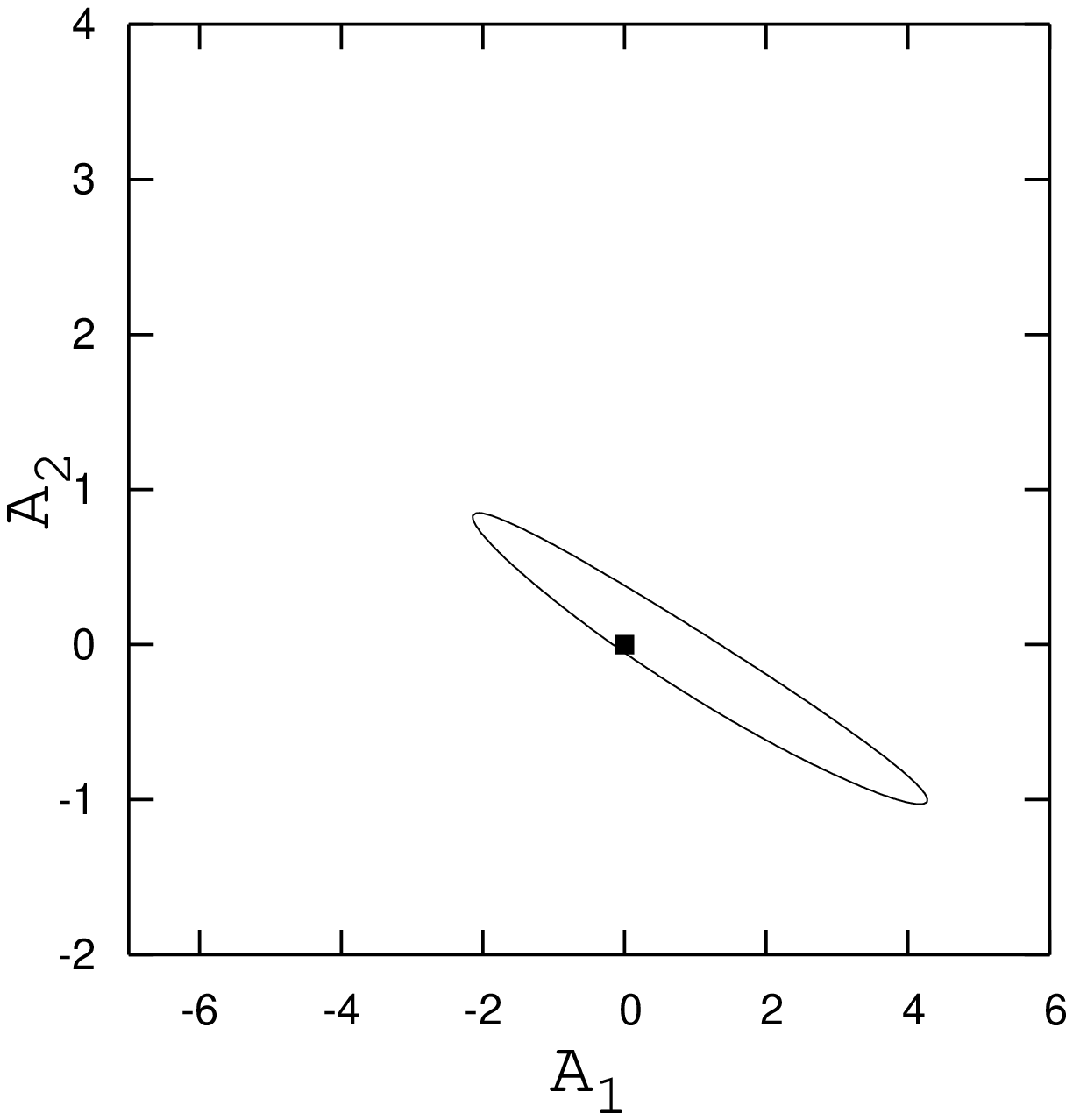} \\
\epsfxsize=2.5in
\epsffile{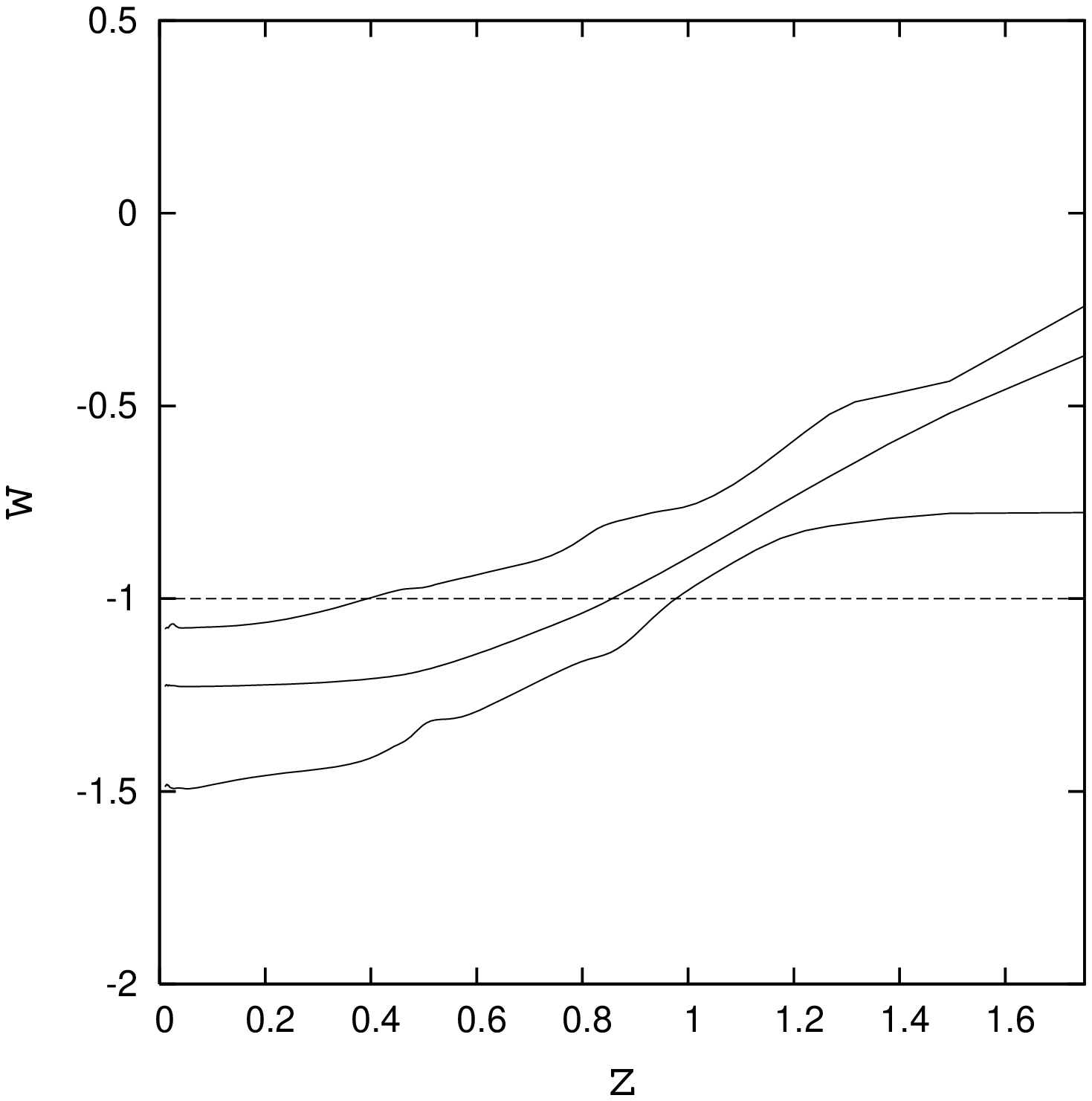} &  
\epsfxsize=2.5in
\epsffile{w_cmb_sdss_om=28.ps} &
\epsfxsize=2.5in
\epsffile{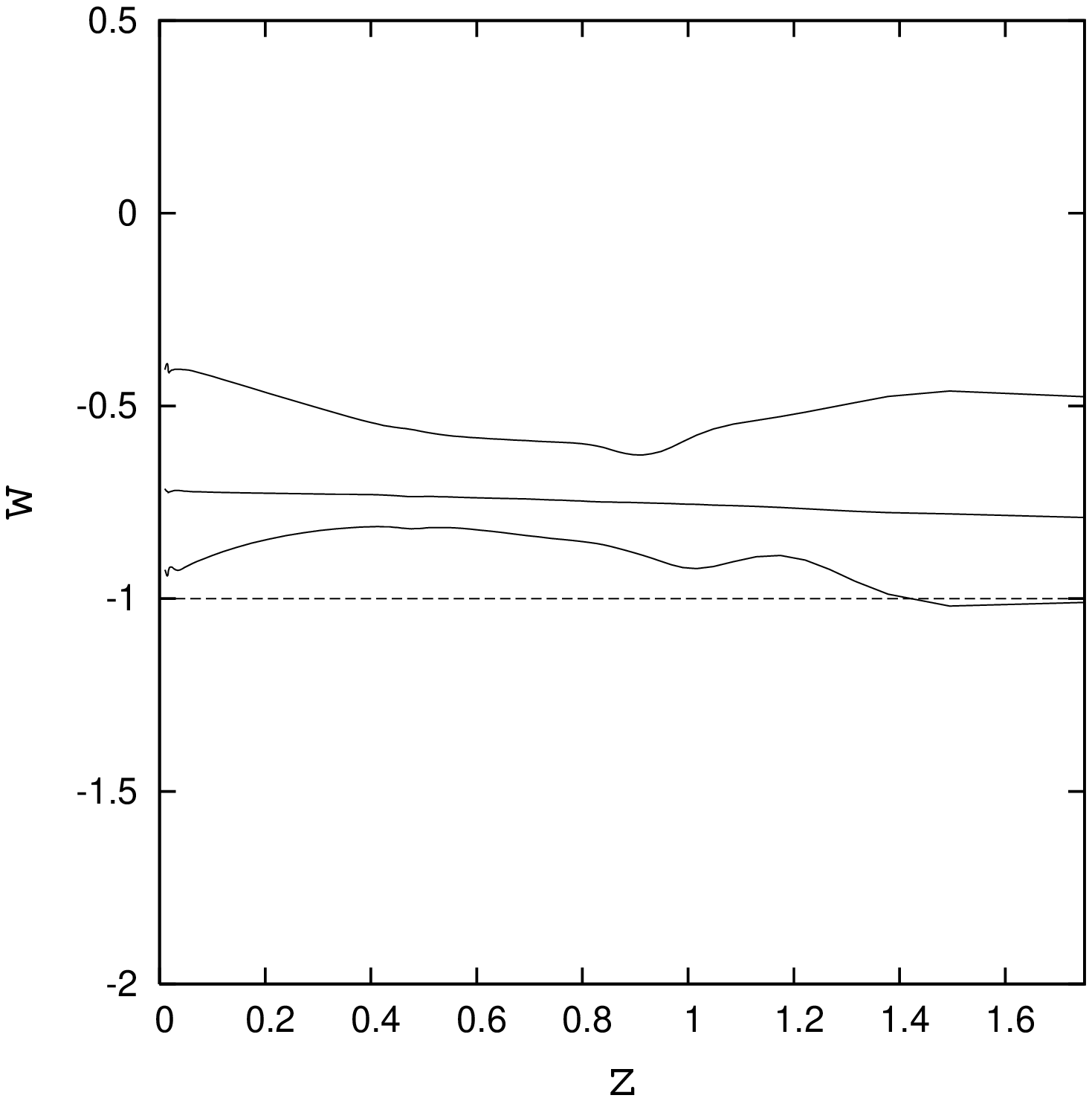} \\
\end{array}$
\end{center}
\vspace{0.0cm}
\caption{\small 
$2 \sigma$ confidence levels for CMB+BAO using different
marginalizations over $\om$. The upper and lower left hand panels
respectively show the confidence levels in $A_1-A_2$ and the $2\sigma$
variation of $w$ over redshift for $\om=0.25\pm0.03$. The middle
panels show the same for $\om=0.28\pm0.03$ and the right-hand panels
are results marginalized over $\om=0.31\pm0.03$.  The black dots in
the upper panels and the dashed lines in the lower panels represent
$\l$CDM. }
\label{fig:cmb_sdss_om}
\end{figure*}

Next we examine our results for the CMB+BAO data. In
figure~\ref{fig:cmb_sdss} we show results for this dataset for which
the observations are at redshifts $z=0.35$ and $z=1089$. We present
our results in panels (b), (c) and (d) over the redshift range of
$z=0-1.7$ in order to be able to compare with our earlier results
obtained using supernova data. The results for a matter density
marginalization of $\om=0.28\pm0.03$ are reasonably consistent with
$\l$CDM, the confidence levels for $A_1-A_2$ are somewhat tighter than
those seen for the SNe results (Comparing panel (a) of
figures~\ref{fig:gold},~\ref{fig:snls},~\ref{fig:cmb_sdss}). However,
unlike the SNe data, the results from CMB+BAO seem to depend much more
strongly on the chosen value of the matter density. In
figure~\ref{fig:cmb_sdss_om} we show the difference in the results for
three different marginalizations of the matter density. In the left
hand panels, we marginalize over $\om=0.25\pm0.03$, and the data then
favours evolving DE which crosses the so-called phantom divide at $w_0
= -1$.  In the middle panels, we marginalize over $\om=0.28\pm0.03$,
the result is a mildly evolving DE model consistent with $\l$CDM. The
panels to the right show results marginalized over a higher value of
the matter density $\om=0.31\pm0.03$, which suggests a DE model with
practically constant $w > -1$. These results are supported by an
analysis of $w$-probe shown in table \ref{tab:wprobe_om} which
demonstrates that the redshift dependence of the mean equation of
state ${\bar w}$ is sensitive to the value of $\om$. Thus, unlike the
SNe data, even a very slight difference in $\om$ can lead to a
significant difference in the degree of evolution of DE for the
CMB+BAO dataset. Note that the generic trend of $\Omega_m$ dependence
is similar to that recently found in \cite{NP06} where all data (the
old Gold+HST sample, the SNLS dataset, WMAP, BAO and others) were
analyzed using the Chevallier-Polarski-Linder parametrization of the
DE equation of state \cite{polar,linder}.

\begin{figure*} 
\centering
\begin{center}
$\begin{array}{@{\hspace{-1.0in}}c@{\hspace{0.0in}}c}
\multicolumn{1}{l}{\mbox{}} &
\multicolumn{1}{l}{\mbox{}} \\ [-0.20in]
\epsfxsize=3.8in
\epsffile{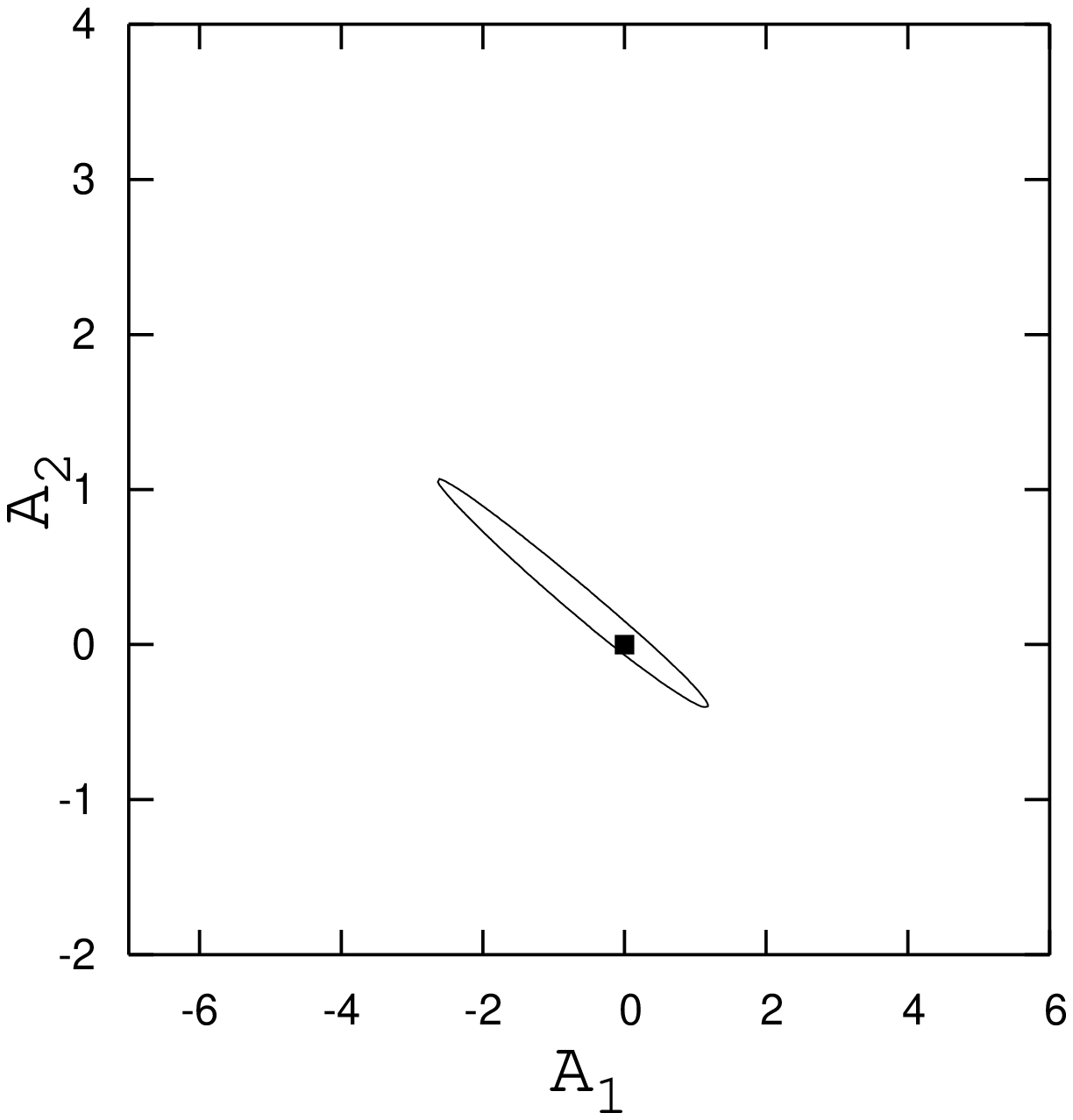} &  
\epsfxsize=3.in
\epsffile{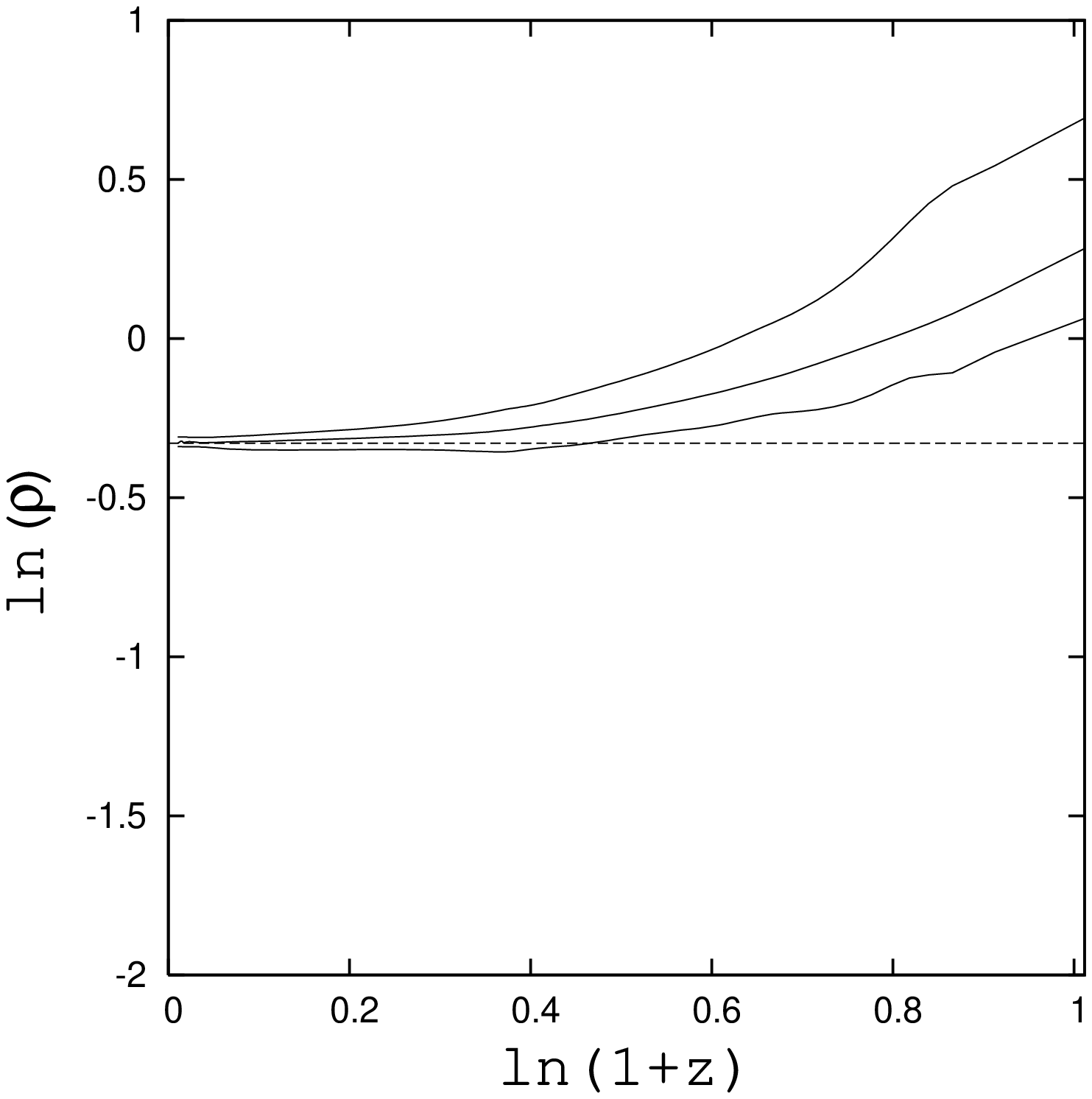} \\
\epsfxsize=3.in
\epsffile{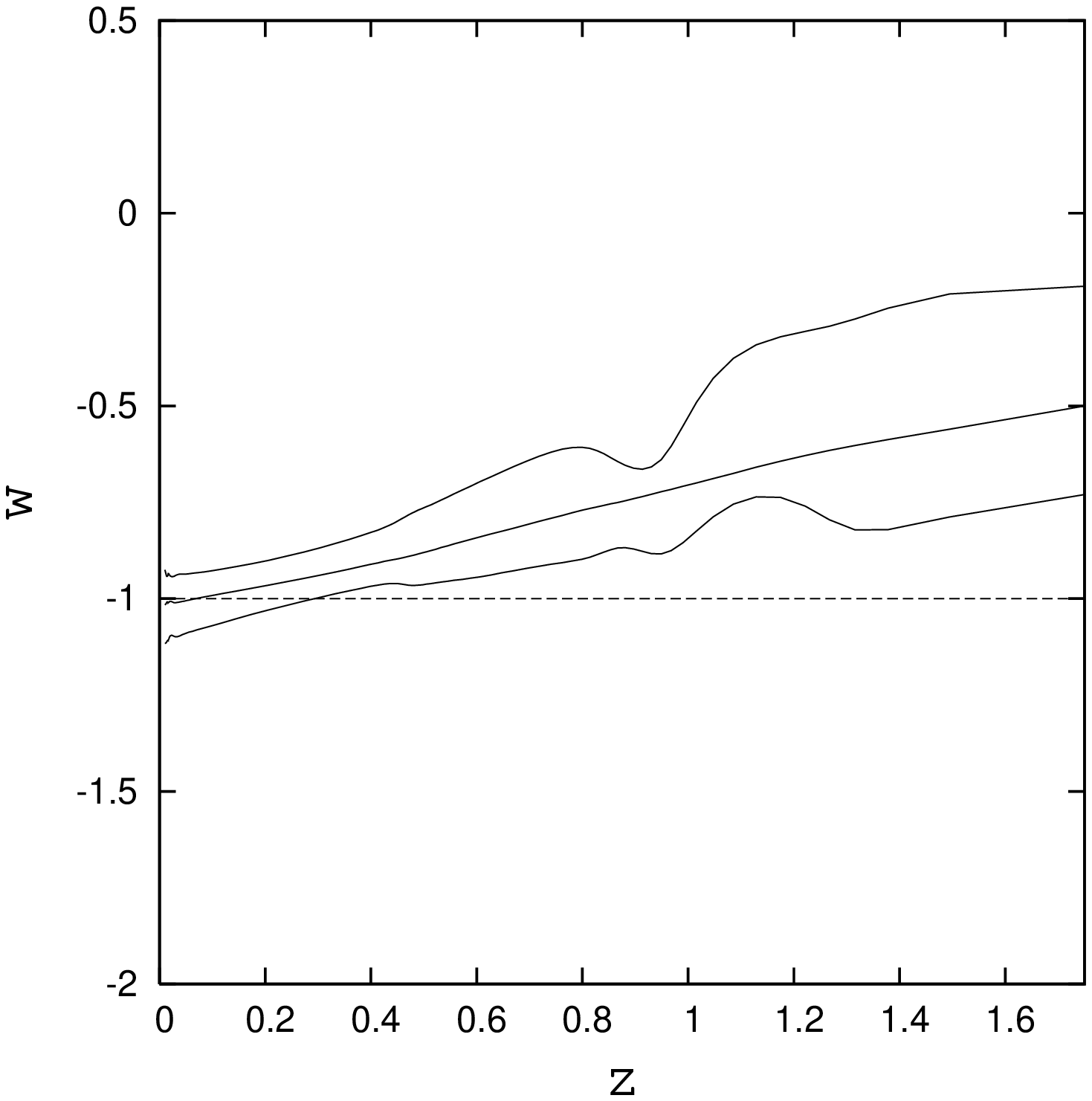} &
\epsfxsize=3.in
\epsffile{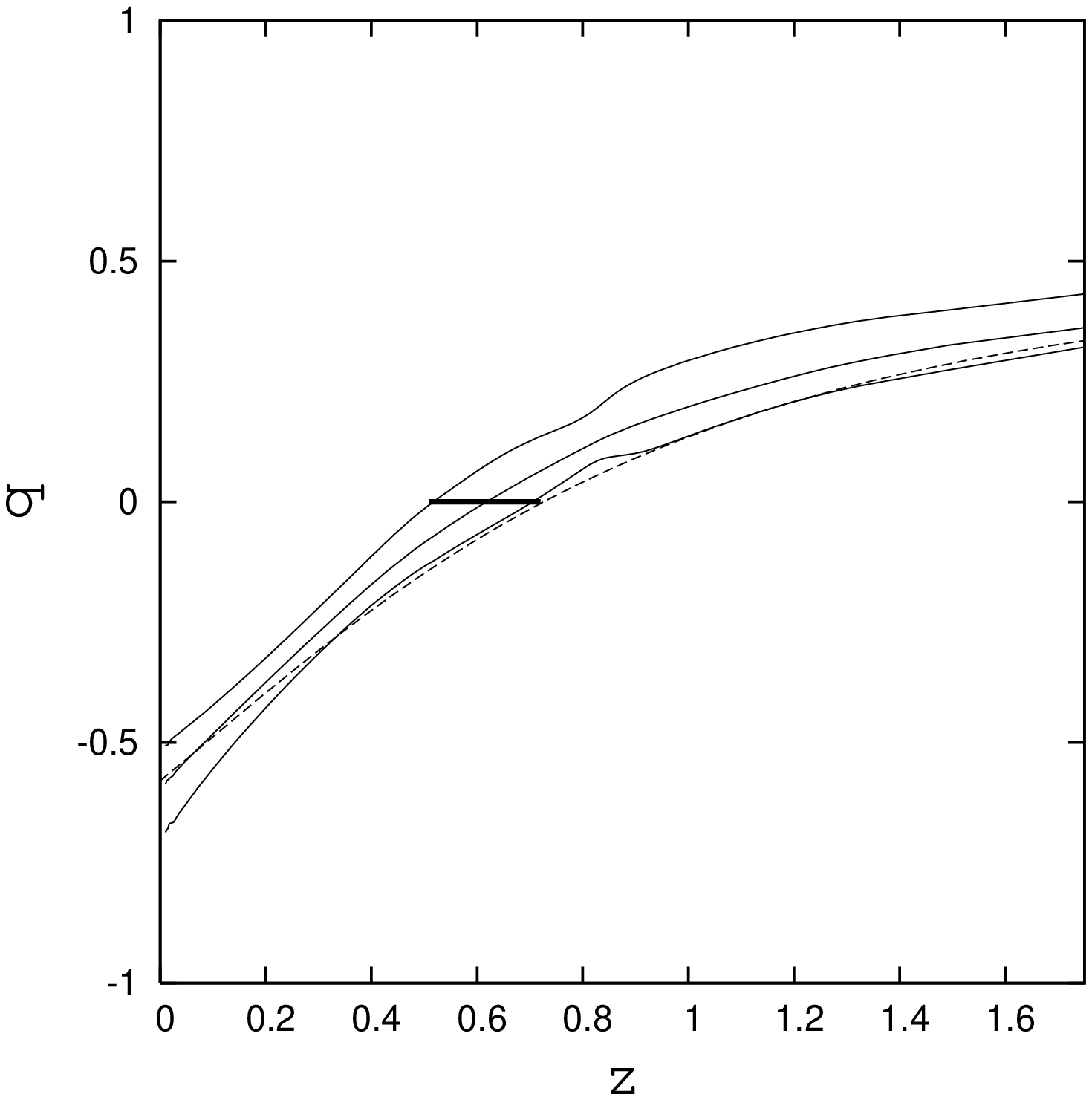} \\
\end{array}$
\end{center}
\vspace{0.0cm}
\caption{\small
$2\sigma$ confidence levels for the Gold+CMB+BAO dataset using
$\om=0.28\pm0.03$. The upper left hand panel shows the confidence
levels in $A_1-A_2$, with the black dot representing $\l$CDM. The
upper right hand panel shows the logarithmic $2\sigma$ variation of
the DE density in terms of redshift. The dashed line
represents $\l$CDM. The lower left and right hand panels represent the
variation of the equation of state and deceleration parameter
respectively. The dashed lines in both panels represent $\l$CDM. The thick
solid line in the lower right hand panel shows the acceleration epoch,
\ie the redshift at which the universe started accelerating.}
\label{fig:cmb_sdss_gold}
\end{figure*}

\begin{figure*} 
\centering
\begin{center}
$\begin{array}{@{\hspace{-1.0in}}c@{\hspace{0.0in}}c}
\multicolumn{1}{l}{\mbox{}} &
\multicolumn{1}{l}{\mbox{}} \\ [-0.20in]
\epsfxsize=3.8in
\epsffile{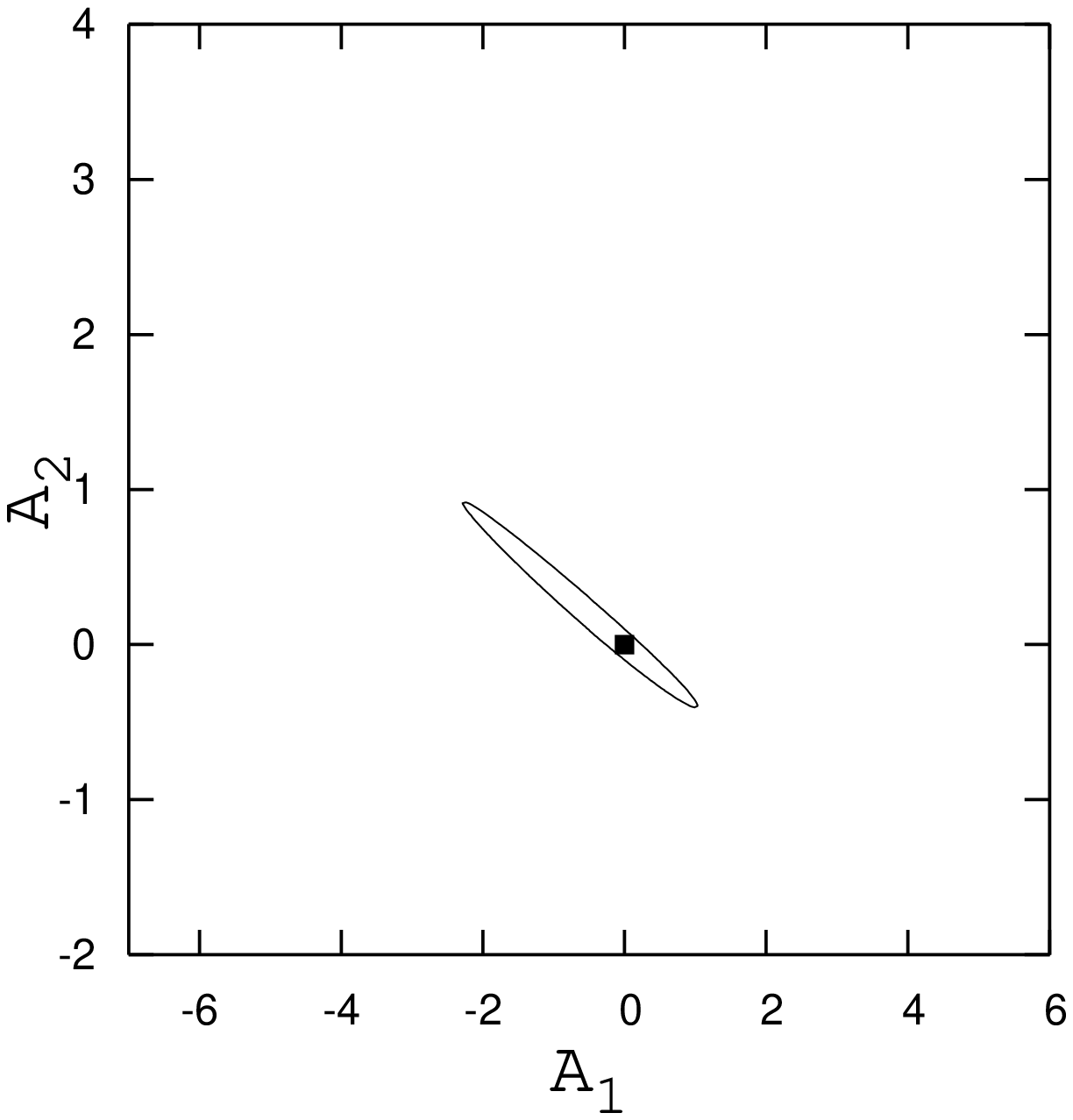} &  
\epsfxsize=3.in
\epsffile{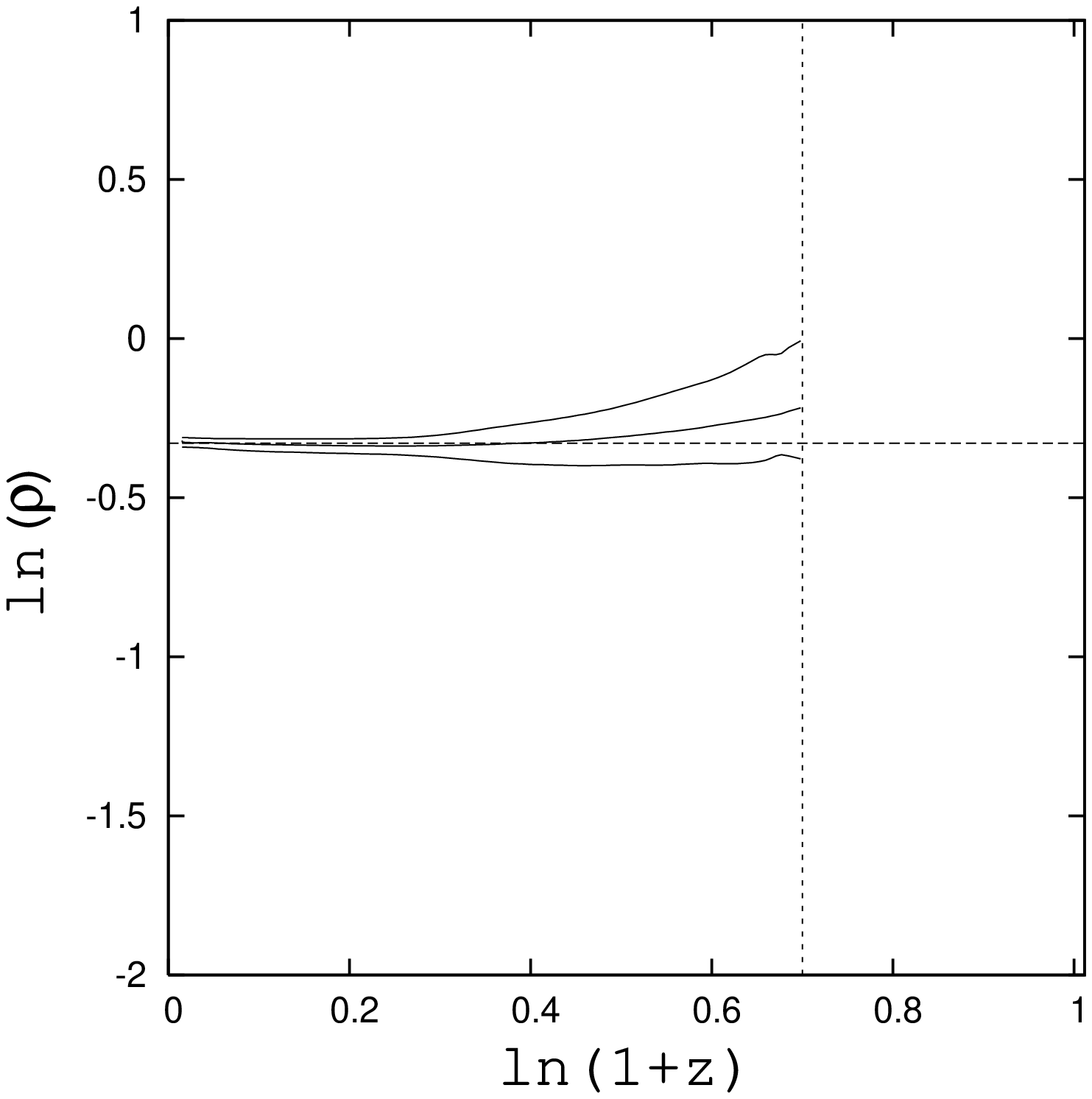} \\
\epsfxsize=3.in
\epsffile{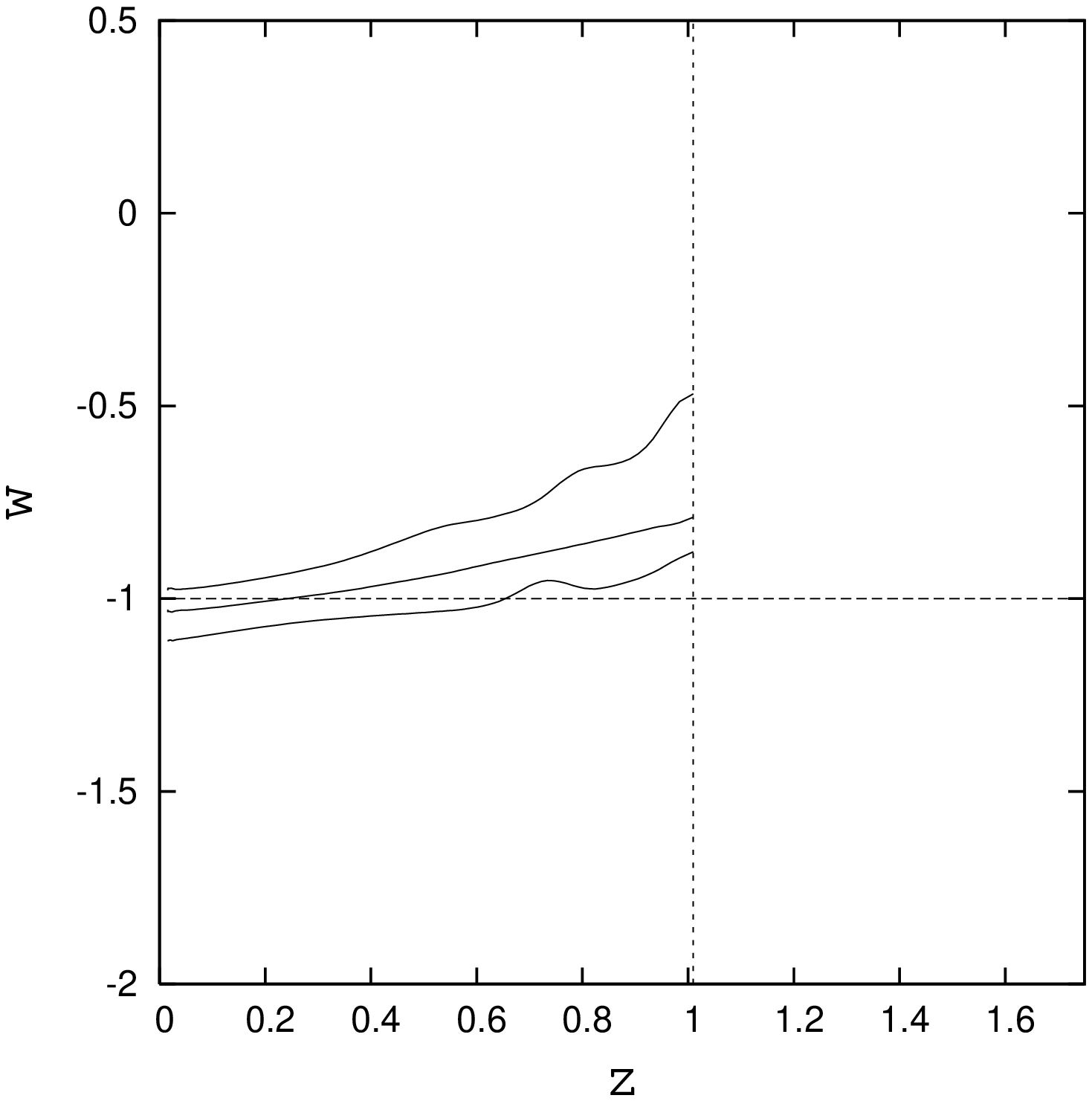} &
\epsfxsize=3.in
\epsffile{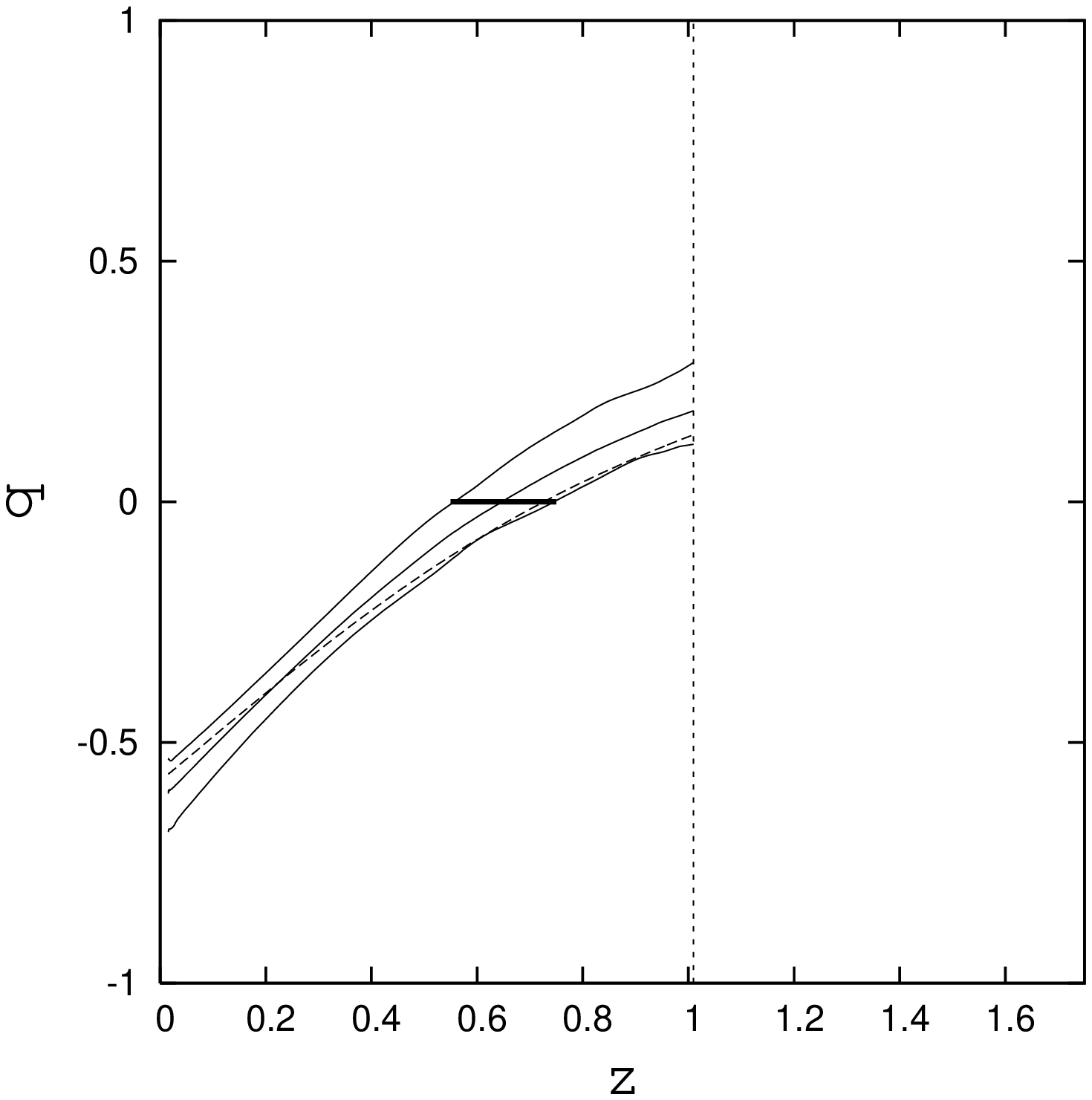} \\
\end{array}$
\end{center}
\vspace{0.0cm}
\caption{\small 
$2\sigma$ confidence levels for the SNLS+CMB+BAO dataset using
$\om=0.28\pm0.03$. The upper left hand panel shows the confidence
levels in $A_1-A_2$, with the black dot representing $\l$CDM. The
upper right hand panel shows the logarithmic $2\sigma$ variation of
the DE density in terms of redshift. The dashed line
represents $\l$CDM. The lower left and right hand panels represent the
variation of the equation of state and deceleration parameter
respectively. The dashed lines in both panels represent $\l$CDM. The thick
solid line in the lower right hand panel shows the acceleration epoch,
\ie the redshift at which the universe started accelerating. Results
are shown upto redshift $z=1.01$}
\label{fig:cmb_sdss_snls}
\end{figure*}

Finally we combine the SNe and CMB+BAO data. We use the
marginalization $\om=0.28\pm0.03$ for the entire dataset. The results
are shown for Gold+HST data in figure~\ref{fig:cmb_sdss_gold}. We see
that the parameter space is strongly constrained by the three sets of
data,the cosmological constant is consistent with the data, so are
mildly evolving DE models. The results from SNLS are shown in
figure~\ref{fig:cmb_sdss_snls}. We see that the results from both sets
of SNe data are strikingly similar, this is because the results are
very strongly constrained by the CMB+BAO data. Whether the CMB+BAO
data should be used in conjunction with the SNe data, given the strong
matter density dependence of the former, is however rather
questionable at this point.

\section{Conclusions}
\label{sec:concl}

In this paper, we have explored the reconstruction of DE using the
most recent and complementary datasets available to us at
present. First, we reconstructed DE using two supernova datasets-- the
recently released Gold+HST dataset and the SNLS two-year dataset. We
find that the results for the two are slightly inconsistent. The
Gold+HST dataset appears to favour an evolving model of DE with $w_0 <
-1$ at present and $w \sim 0$ at $z \sim 1$ over $\l$CDM for its
best-fit, however $\l$CDM is still consistent at $2\sigma$. For SNLS
data, the results are more consistent with $\l$CDM. It should be
emphasized here that we are speaking about DE properties averaged over
a redshift interval $\sim 0.4$ or more, see e.g. Tables I, II. Note
also that, as follows from these tables and Figs. 2-7, significantly
phantom behaviour of DE ($w\lesssim -1.1$), if it occurs at all, is
possible for moderately low redshifts $z < 0.4$ only and is favoured
by the Gold+HST dataset, but not by the SNLS data.

This discrepancy may be a result of the different light-curve
standardization techniques used by the two teams or due to systematic
bias.  Also, this could be due to the fact that the low redshift and
high redshift SNe are obtained by different surveys. Future surveys
such as SNAP \cite{snap}, JEDI \cite{jedi} and DUNE \cite{dune} should
have a better handle on the systematic errors and such discrepancies
should disappear.

We also attempt to obtain information on DE using datasets
complementary to the supernova sample. To this end, we use information
on the shift parameter $R$ from the 3-year WMAP data and the quantity
$A$ for the baryon acoustic oscillation peak from SDSS. The results
are consistent with $\l$CDM but do not rule out weakly time dependent
DE. These datasets appear to be rather sensitive to the value chosen
for the present matter density $\om$. As a result, it is difficult to
reach firm conclusion on the nature of DE from these data until strong
model-independent constraints on $\om$ are obtained. Thus, even with
the most recent data, the fundamental question if DE reduces to
$\Lambda$ or not, still remains open. On the other hand, possible
deviation of DE properties from those of $\Lambda$ is gradually
becoming more and more restricted.

While this work was being finalised the paper \cite{gong_wang}
appeared containing results which are in broad agreement with ours in
areas of overlap.

\section*{Acknowledgements}

AS was partially supported by the Russian Foundation for Basic Research, 
grant 05-02-17450, and by the Research Program ``Astronomy" of the Russian 
Academy of Sciences. He also thanks the Centre Emile Borel, Institut Henri 
Poincar\'e, Paris for the hospitality and the CNRS for partial support 
during the period when this project was finished.

\newpage

\begin{table}
\begin{center} 
\caption{\scriptsize The reconstructed $w$-probe for the two different SNe datasets, Gold+HST and SNLS, using $\om=0.28\pm0.03$}
\begin{tabular}{cccc}
&&$\bar{w}$&\\
Dataset&$~~\Delta z=0-0.414 ~~$&$ ~~\Delta z=0.414-1 ~~$&$ ~~\Delta z=1-1.755 ~~~$\\
\hline
Gold+HST&$-1.160^{+0.089}_{-0.070}$&$-0.226^{+0.319}_{-0.259}$&$0.268^{+0.073}_{-0.041}$\\
SNLS&$-1.037^{+0.069}_{-0.072}$&$-0.985^{+0.428}_{-0.296}$&\\
\end{tabular}\label{tab:wprobe}
\end{center}
\end{table}

\begin{table}
\begin{center} 
\caption{\scriptsize The reconstructed $w$-probe for the CMB+BAO dataset using three different marginalizations over $\om$}
\begin{tabular}{cccc}
&&$\bar{w}$&\\
$\om$&$\Delta z=0-0.414 ~~$&$ ~~\Delta z=0.414-1 ~~$&$ ~~\Delta z=1-1.755 ~~$\\
\hline
$0.25\pm0.03$&$-1.195^{+0.089}_{-0.070}$&$-0.948^{+0.319}_{-0.259}$&$-0.493^{+0.071}_{-0.041}$\\
$0.28\pm0.03$&$-0.951^{+0.078}_{-0.072}$&$-0.779^{+0.428}_{-0.296}$&$-0.629^{+0.432}_{-0.296}$\\
$0.31\pm0.03$&$-0.772^{+0.091}_{-0.079}$&$-0.819^{+0.394}_{-0.225}$&$-0.852^{+0.521}_{-0.535}$\\
\end{tabular}\label{tab:wprobe_om}
\end{center}
\end{table}

\newpage
\newpage

\end{document}